\pdfoutput=1

\documentclass[10pt,journal,compsoc]{IEEEtran}
\ifCLASSOPTIONcompsoc
  \usepackage[nocompress]{cite}
\else
  \usepackage{cite}
\fi
\ifCLASSINFOpdf
   \usepackage[pdftex]{graphicx}
   \usepackage{xcolor}
\else
\fi
\usepackage{amsmath}
\usepackage{amsfonts}
\usepackage{algorithmic}
\usepackage{subfigure}
\usepackage{multirow, booktabs}
\usepackage{graphicx}
\usepackage{pifont}
\usepackage{tikz}

\begin{document}
%
\title{Privacy-Preserving Aggregation in Federated Learning: A Survey}
%
%
%
%

\author{Ziyao~Liu,
        Jiale~Guo,
        Wenzhuo~Yang,
        Jiani~Fan,
        Kwok-Yan~Lam,~\IEEEmembership{Senior Member,~IEEE,}
        and~Jun~Zhao,~\IEEEmembership{Member,~IEEE}
\IEEEcompsocitemizethanks{\IEEEcompsocthanksitem Ziyao Liu, Jiale Guo, Wenzhuo Yang, Jiani Fan, Kwok-Yan Lam, and Jun Zhao are with the School of Computer Science and Engineering, Nanyang Technological University, Singapore, 50 Nanyang Ave, 639798.\protect\\
E-mail: \{ziyao002, jiale001, wenzhuo001, jiani001\} @e.ntu.edu.sg, \{kwokyan.lam, junzhao\} @ntu.edu.sg}
\thanks{Manuscript received April 19, 2005; revised August 26, 2015.}}

%
%

\markboth{Journal of \LaTeX\ Class Files,~Vol.~14, No.~8, August~2015}%
{Shell \MakeLowercase{\textit{et al.}}: Bare Demo of IEEEtran.cls for Computer Society Journals}
%



\IEEEtitleabstractindextext{%
\begin{abstract}
Over the recent years, with the increasing adoption of Federated Learning (FL) algorithms and growing concerns over personal data privacy, Privacy-Preserving Federated Learning (PPFL) has attracted tremendous attention from both academia and industry. Practical PPFL typically allows multiple participants to individually train their machine learning models, which are then aggregated to construct a global model in a privacy-preserving manner. As such, Privacy-Preserving Aggregation (PPAgg) as the key protocol in PPFL has received substantial research interest. This survey aims to fill the gap between a large number of studies on PPFL, where PPAgg is adopted to provide a privacy guarantee, and the lack of a comprehensive survey on the PPAgg protocols applied in FL systems. In this survey, we review the PPAgg protocols proposed to address privacy and security issues in FL systems. The focus is placed on the construction of PPAgg protocols with an extensive analysis of the advantages and disadvantages of these selected PPAgg protocols and solutions. Additionally, we discuss the open-source FL frameworks that support PPAgg. Finally, we highlight important challenges and future research directions for applying PPAgg to FL systems and the combination of PPAgg with other technologies for further security improvement.

\end{abstract}


}

\maketitle

\IEEEdisplaynontitleabstractindextext

%
\IEEEpeerreviewmaketitle

\IEEEraisesectionheading{\section{Introduction}
\label{sec:introduction}}

%
%
%
%
\IEEEPARstart{O}{ver} the recent years, with the increasing adoption of machine learning (ML) algorithms and growing concern of data privacy, the scenario where different data owners, e.g., mobile devices or cloud servers, jointly solve a machine learning problem, i.e., train an ML model, while preserving their data privacy has attracted tremendous attention from both academia and industry. In this connection, federated learning (FL) \cite{yang2019federated} is proposed to achieve privacy-enhanced distributed machine learning schemes, and has been applied to a wide range of scenarios such as Internet of Things (IoT) \cite{huang2021starfl,yang2022lead,li2021federated}, healthcare \cite{liu2022contribution,ju2020federated,chen2021fl,chen2020dealing}, computer vision \cite{liu2020fedvision,liu2021federated,luo2019real}, and recommendation \cite{tan2020federated,yang2020federated}. A standard FL system typically enables different participants, i.e., data owners, to individually train an ML model using their local data, which are then aggregated by a central server to construct a global FL model. However, as pointed out in \cite{zhu2019deep}, with only a small portion of the user's model, an attacker, e.g., a malicious central server, can easily reconstruct the user's data with pixel-wise accuracy for images and token-wise matching for texts. To mitigate such so-called ``deep leakage from gradients", Privacy-Preserving Technique (PPT) such as Homomorphic Encryption (HE) \cite{gentry2009fully}, Multi-Party Computation (MPC) \cite{yao1982protocols}, Differential Privacy (DP) \cite{dwork2014algorithmic}, and infrastructures such as blockchain \cite{nakamoto2008bitcoin} and Trusted Execution Environment (TEE) \cite{tee} have been proposed to enhance FL systems by aggregating the users' locally trained models in a privacy-preserving manner. As such, Privacy-Preserving Aggregation (PPAgg) as the key protocol in Privacy-Preserving Federated Learning (PPFL) has received substantial research interest.

In general, one can enhance PPFL by constructing PPAgg protocols that are widely adopted in standard distributed Privacy-Preserving Machine Learning (PPML). However, compared to PPML, PPFL further considers heterogeneous participants of, e.g., different computational power and bandwidth, and more complicated threat models regarding privacy and security \cite{kairouz2021advances}. For example, in a cross-device FL setting, participants are usually resource-constrained mobile devices that may drop out of the system at any time (see Section \ref{sec:fl} for more details). This requires the PPAgg protocol to provide both cost-effective execution and dropout resilience. Meanwhile, the security and privacy issues in PPFL systems may come from insiders, e.g., FL participants, or outsiders, e.g., simulated dummy participants, from a single adversary, e.g., the central server, or multiple adversaries, e.g., several colluding participants. Besides, adversaries can be considered to be semi-honest, i.e., try to learn the private information of honest participants without deviating from the FL protocol, or active malicious, i.e., try to learn the private information of other honest participants by deviating arbitrarily from the FL protocol, e.g., by manipulating messages. Therefore, specific designs of PPAgg protocols are required to achieve PPFL in different application scenarios.

\textbf{Comparison with other surveys.}
Currently, few existing surveys on PPFL perceive the construction and organization of PPFL from the perspective of privacy-preserving aggregation protocols. In particular, the surveys in \cite{yang2019federated,kairouz2021advances,yang2021toward} give a comprehensive introduction to federated learning. The surveys in \cite{lyu2020privacy,lyu2020threats} extensively analyze the privacy and security threats to FL systems with discussions on possible attacks and defenses. The survey in \cite{briggs2021review,fang2021privacy} presents the PPFL applications to the Internet of Things, and the surveys in \cite{xia2021survey,lim2020federated} discuss the integration of PPFL and edge computing. Several research papers such as \cite{fereidooni2021safelearn,mondal2022beas,liu2022efficient} have surveyed some PPAgg protocols in FL. However, they do not provide extensive discussion regarding different constructions and threat models. To the best of our knowledge, there is no survey specifically discussing the aggregation protocols, as a key privacy-preserving technique, adopted in PPFL systems. This motivates us to deliver the survey with a comprehensive literature review on the construction of PPAgg protocols in PPFL with a discussion on their application scenarios. We note that many studies focus on optimizing the performance and efficiency of standard FL training. However, this survey concentrates on FL systems from a privacy and security perspective, hence they are out of the scope of this paper, and interested readers can refer to \cite{kairouz2021advances,jin2020towards,tan2021towards} for the surveys on state-of-the-art FL training algorithms. For convenience, the related works in this survey are classified based on their main technique used to guarantee privacy, as one PPAgg protocol may involve several supported privacy-preserving techniques to provide different properties. For example, SecAgg \cite{bonawitz2017practical}, which adopts both masking and secret sharing technique, is classified as a masking-based aggregation in this survey since the masking technique is deployed to protect the users' model privacy while secret sharing mainly provides the dropout-resilience. These major classifications consist of (i) masking-based aggregation, (ii) HE-based aggregation, (iii) MPC-based aggregation, (iv) DP-based aggregation, (v) blockchain-based aggregation, and (vi) TEE-based aggregation.

\textbf{Organisation of the paper.}
The rest of this paper is organized as follows. Section \ref{sec:fl} describes the general architecture of and privacy threats to federated learning systems. Section \ref{sec:sup-tools} presents the fundamentals of supporting tools that are commonly used for privacy-preserving aggregation. Section \ref{sec:ppa-protocols} reviews different constructions of PPAgg protocols in federated learning, followed by the discussions on open-source FL frameworks that support PPAgg in Section \ref{sec:fl-frameworks}. Section \ref{sec:challenges} outlines challenges and future research directions. Section \ref{sec:conclusions} summarizes and concludes the paper.

%
%

\section{Overview and Fundamentals of Federated Learning}
\label{sec:fl}
In this section, we will give an overview of the federated learning on its concepts, data organization, working mechanism, and privacy threats to FL systems.

\subsection{Overview of Federated Learning}

A federated learning scheme typically enables different participants, i.e., data owners, to individually train an ML model using their local data, which are then aggregated with the coordination of a central server to construct a global FL model. The FL participants can be divided into two classes, i.e., (i) a set of $n$ users $\mathcal{U}=\{u_1,u_2,\dots,u_n\}$ that each user $u_i \in \mathcal{U}$ has a local dataset $\mathcal{D}_i$, and (ii) a central server $S$.

\begin{figure}[htbp]
    \centering
    \includegraphics[width=0.7\linewidth]{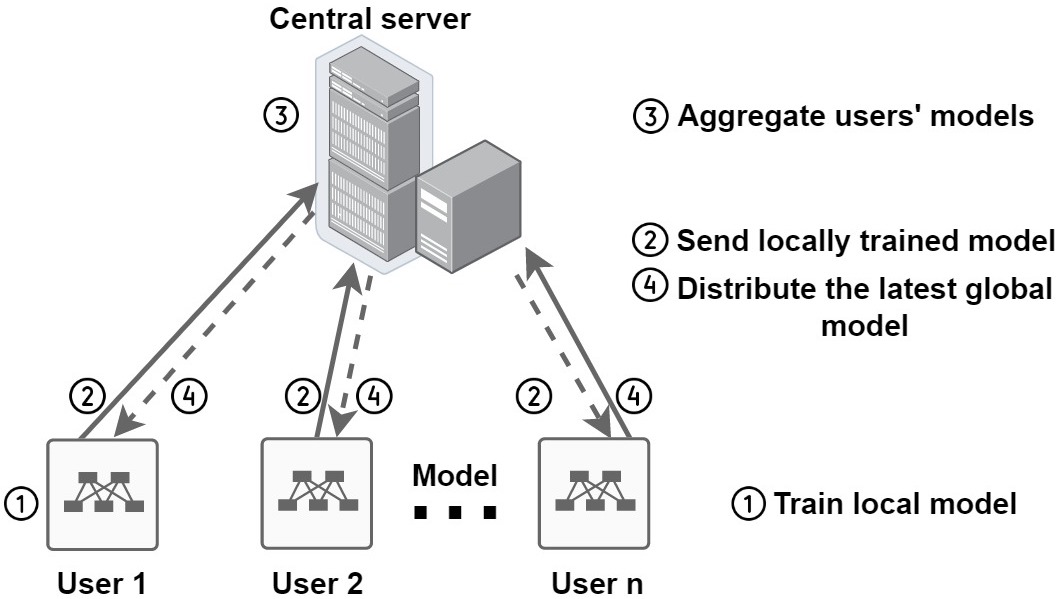}
    \caption{A typical workflow of the training process in FL systems.}
    \label{fig:fl-workflow}
\end{figure}

As shown in Figure \ref{fig:fl-workflow}, a typical FL scheme works by repeating the following steps until training is stopped. (i) Local model training: each FL user $u_i$ trains its model $\mathcal{M}_i$ using the local dataset $\mathcal{D}_i$. (ii) Model uploading: each FL user $u_i$ uploads its locally trained model $\mathcal{M}_i$ to the central server $S$. (iii) Model aggregation: the central server $S$ collects and aggregates users' models to update the global model $\mathcal{M}$. (iv) Model updating: the central server $S$ updates the global model $\mathcal{M}$ and distributes it to all FL users.


Furthermore, as mentioned earlier, FL systems usually involve heterogeneity with respect to the data format, computational power, bandwidth, etc. Therefore, specific designs of PPAgg protocols are required to achieve PPFL in different settings. Here we keep the consistency of the classification of FL settings from \cite{kairouz2021advances}, i.e., Cross-Silo setting and Cross-Device setting, as described in Table \ref{tab:fl-setting}. 

\begin{table}[thb]
\scriptsize
\centering
\setlength{\abovecaptionskip}{-0.05cm}
\caption{An adaptive classification of FL settings from \cite{kairouz2021advances}.}
\label{tab:fl-setting}
\begin{tabular}{|c|l|l|} 
\hline
& \multicolumn{1}{c|}{\textbf{Cross-Silo}}                                 & \multicolumn{1}{c|}{\textbf{Cross-Device}}                                         \\ 
\hline
\textbf{Participants}                                                         & \begin{tabular}[c]{@{}l@{}}Several different\\organizations\end{tabular} & \begin{tabular}[c]{@{}l@{}}A large number of \\mobile or IoT devices\end{tabular}  \\ 
\hline
\begin{tabular}[c]{@{}c@{}}\textbf{Distribution}\\\textbf{scale}\end{tabular} & Typically 2-100 users                                                    & Up to $10^{10}$ users                                                                \\ 
\hline
\begin{tabular}[c]{@{}c@{}}\textbf{Primary}\\\textbf{bottleneck}\end{tabular} & \begin{tabular}[c]{@{}l@{}}Computation or \\communication\end{tabular}   & \begin{tabular}[c]{@{}l@{}}Communication due to \\slow connections\end{tabular}    \\ 
\hline
\begin{tabular}[c]{@{}c@{}}\textbf{User }\\\textbf{reliability}\end{tabular}  & Relatively few failures                                                  & Highly unreliable                                                                  \\ 
\hline
\begin{tabular}[c]{@{}c@{}}\textbf{User}\\\textbf{statefulness}\end{tabular}  & \begin{tabular}[c]{@{}l@{}}Online from round \\to round\end{tabular}     & May drop out at anytime                                                            \\
\hline
\end{tabular}
\end{table}
\vspace{-0.4cm}

\subsection{Privacy Threats to Federated Learning}

As discussed earlier, privacy threats in FL systems can lead to different information leakage and come from both insiders, e.g., FL participants, and outsiders, e.g., eavesdroppers. However, the capability of insider adversaries is generally stronger than that of outsider adversaries, as one can adopt cryptographic tools to, e.g., achieve secure communication channels and verify the identities of outsiders, to mitigate the privacy issue caused by outsider attacks. Therefore, our discussion of privacy threats in FL will focus primarily on insider privacy leakages. We should note that in this survey, we focus on PPAgg protocols that protect the privacy of honest FL participants' inputs. The protocols that aim to guarantee security against non-privacy attacks, e.g., backdoor attacks or poisoning attacks \cite{bagdasaryan2020backdoor}, fall into Byzantine-robust aggregation \cite{blanchard2017machine,yin2018byzantine,CaoF0G21} which are out of the scope of this paper. However, in Section \ref{sec:discussion}, we will discuss the potential integration of them with PPAgg protocols for further security improvement. The privacy threats in FL can be categorized into the following two general forms.

\begin{itemize}
    \item Privacy threats to users' models: it is proved that local data of an individual FL participant could be revealed through a small portion of its locally trained model \cite{zhu2019deep}, which directly breaks the basic privacy guarantee of standard federated learning. Therefore, a large number of PPAgg protocols focus on dealing with such privacy leakage.
    \item Privacy threats to the global model: standard FL schemes assume that the global models are over plaintext. However, the privacy of the global models is considered in some FL application scenarios. Therefore, PPAgg protocols are required to provide the privacy guarantee of both users' models and the global model.
\end{itemize}

Furthermore, privacy leakages may come from a single adversary or multiple adversaries, which can take one of the following two general forms.

\begin{itemize}
    \item Single adversary: a single, non-colluding participant, which can be an FL user, the central server, or the third party. Note that the central server usually has a stronger capability than a single FL user.
    \item Colluding adversaries: collusion may happen with or without the central server. Note that colluding FL users with the central server and more adversaries usually lead to a greater risk to privacy leakages.
\end{itemize}

Last but not least, the capability of adversaries should be taken into account, which can be classified into the following three general forms.

\begin{itemize}
    \item Honest: follow the protocol honestly.
    \item Passive malicious (honest-but-curious or semi-honest): try to learn the private information of honest participants without deviating from the protocol.
    \item Active malicious (malicious when the context is clear): can deviate from the protocol at any time by any means, e.g., manipulating identity or sending fraudulent messages to others.
\end{itemize}

\section{Techniques for Privacy-Preserving Aggregation}
\label{sec:sup-tools}

In this section, we give an overview of the supporting tools to construct privacy-preserving aggregation protocols in FL systems, including some privacy-preserving techniques such as one-time pad, homomorphic encryption, secure multi-party computation, differential privacy, and infrastructures such as blockchain and trusted execution environment.

\subsection{One-time Pad}
\label{sec:one-time pad}
In cryptography, the One-Time Pad (OTP) is a technique in which the sender randomly generates a private key that will be used only once to encrypt a message. For decryption, the receiver needs to use a matching OTP as the key. In such a way, the randomness of OTP guarantees that each encryption of message is unique and has no relation to any other encryption, and thus there is no way to break the messages encrypted by OTP. Therefore, OTP crypto-systems provide provably unconditional security \cite{katz2020introduction}.

In specific, OTP can be adopted to encrypt a message in an additive or multiplicative manner. For example, it can be easily proved that by adding random generated OTP $r$ to a message $x$ in a finite field $\mathbb{F}_p$ to obtain the encrypted message $y$, i.e., $y=x+r \bmod{p}$, the message $x$ is perfectly masked by $r$. In other words, there is no way for an attacker to break the code of $y$ unless the $r$ is revealed. Similarly, the multiplicative masking by OTP has the same security guarantee as that of additive masking, provided the message $x \neq 0$, i.e., $y=x\cdot r \bmod{p}$. Following this way, FL participants can mask their models to preserve their privacy. However, keeping the correctness of aggregation on the masked model is not a straightforward task. Therefore, well-designed masking techniques with aggregation protocols are proposed to cancel the masks to get the correct results. We will review those related works in Section \ref{sec:masking-based aggregation}.


Note that OTP-based masking is different from DP-based perturbation. The reason is that OTP provides perfect secrecy but DP still leaks some statistical information of the database. In addition, OTP-based, i.e., masking-based, aggregation usually provides exact results while DP-based aggregation inevitably suffers from noise, hence the degradation of FL model performance. Throughout this paper, for a vector or a model encrypted by OTPs, we call them masked vector or masked model. We should note that unconditional security can be guaranteed only over the finite field. Thus, for computations on fixed-point numbers that are widely used in federated learning systems, one has to first convert those numbers to field elements in order to be compatible with privacy-preserving aggregation protocols.

\subsection{Homomorphic Encryption}
\label{sec:he}

Homomorphic Encryption (HE) is a kind of encryption scheme that allows one to perform function evaluations over encrypted data while preserving the function features and data format. As an example of additive public-key HE scheme with the key pair $(pk,sk)$, for two messages $m_1$ and $m_2$, one can compute $Enc(m_1+m_2,pk)$ using $Enc(m_1,pk)$ and $Enc(m_2,pk)$ without knowing any information about $m_1$ and $m_2$, where $Enc(\cdot)$ denotes the encryption function and $pk$ is the public key. After that, one can obtain $m_1+m_2$ relying on the corresponding decryption function $Dec(\cdot)$ and the secret key $sk$. Note that for simplicity, we sometimes abuse the notation $Enc(\cdot)$ and $Dec(\cdot)$ without using $pk$ and $sk$ when the context is clear.

In general, HE schemes can be categorized according to the number of allowed arithmetic operations on the encrypted data as follows.

\begin{itemize}
    \item Partially Homomorphic Encryption (PHE): allows an unlimited number of operations but with only one type, e.g., addition or multiplication.
    \item Somewhat Homomorphic Encryption (SWHE): allows some types of operations but with a limited number of times, e.g., one multiplication with unlimited number of additions.
    \item Fully Homomorphic Encryption (FHE): allows an unlimited types of arithmetic operations with unlimited number of times.
\end{itemize}

For privacy-preserving aggregation in FL systems, as it involves only one type of arithmetic operation, i.e., addition, the PHE scheme becomes the natural option. For example, Paillier crypto-systems \cite{paillier1999public} are widely adopted in FL to enable addition over encrypted data, hence protecting users' privacy. ElGamal crypto-systems \cite{elgamal1985public} can also be adapted by converting aggregation to product. Furthermore, to protect the privacy of the whole FL workflow, e.g., the global model, FL users need to train their local model based on an encrypted global model, which requires complicated function evaluation over ciphertext. Therefore, FHE schemes are considered to be the only choice. Otherwise, one has to convert all encrypted data to a secretly shared format and leverage MPC protocols to train the ML model. Among FHE schemes, lattice-based CKKS \cite{cheon2017homomorphic,sphinx} is the most popular scheme used in privacy-preserving FL due to the good trade-off with respect to their efficiency and accuracy. Note that PHE schemes are usually more efficient than SWHE, while SWHE schemes are usually more efficient than FHE. Due to the privacy requirements of PPFL systems, SWHE outperforms neither PHE for privacy-preserving aggregation nor FHE for privacy-preserving training. Thus, SWHE is often considered to be the underlying tool to support other high-level cryptographic protocols. For example, leveled BGV (a type of SWHE) is used to construct the generic MPC protocol SPDZ \cite{rotaru2022actively}. Besides, we should note that by sacrificing some efficiency, all HE schemes can be extended to their threshold or multi-key version, e.g., threshold Paillier \cite{nishide2010distributed} and multi-key CKKS \cite{mouchet2020multiparty}. In this case, the secret key is distributed among all participants that are involved in the key generation process. Hence, one has to corrupt more FL participants to break the security compared to those of standard HE schemes, which improves the security.

\subsection{Secure Multi-Party Computation}
\label{sec:mpc}

Secure Multi-Party Computation (MPC or SMPC) broadly encompasses all cryptographic techniques for privacy-preserving function evaluations between multiple parties, including but not limited to homomorphic encryption (HE), Garbled Circuit (GC), Oblivious Transfer (OT), and Secret Sharing Scheme (SSS). Since its general definition in \cite{yao1982protocols}, MPC has moved from pure theoretical interests to practical implementations \cite{bogetoft2009secure,bogdanov2008sharemind}, and has developed many generic frameworks to support secure computation in two-party, e.g., ABY \cite{demmler2015aby}, and in multi-party settings, e.g., SPDZ family \cite{damgaard2012multiparty} and ABY3 \cite{mohassel2018aby3}. Besides, with the development of machine learning during recent years, an efficient MPC scheme supporting privacy-preserving machine learning has attracted tremendous attention from both academia and industry, e.g., \cite{demmler2015aby,mohassel2017secureml,mohassel2018aby3,wagh2019securenn,byali2020flash,chaudhari2019trident}. Note that these generic PPML schemes can be straightforwardly extended to achieve PPFL systems where users share their local data or locally trained models to several participants, e.g., non-colluding servers, that keep online during the whole protocol execution (see Section \ref{sec:mpc-agg}). Since the applications of pure-MPC to FL usually lead to impractical communication overheads, especially for large-scale FL systems with complicated ML models, simple secret sharing schemes are always considered to be integrated with other PPTs to achieve privacy-preserving aggregation or training in FL.


Secret sharing refers to a cryptographic primitive that allows a secret to be distributed and reconstructed among a set of participants. More formally, a $(t,n)$ threshold secret sharing scheme allows one to distribute a secret $s$ to $n$ parties $p_1,p_2,\dots,p_n$ such that only a subset of these parties of which the number is not less than the threshold $t$ can reconstruct the secret $s$, while any subset of parties of which the number is less than the threshold $t$ does not obtain any information about the secret $s$. In specific, additive secret sharing and Shamir secret sharing are two widely-used schemes to construct such MPC protocols. A secret sharing is linear or additive if the reconstruction of the secret from the shares is a linear mapping or additive homomorphic. For example, in an additive secret sharing scheme, the secret $s$ is divided into $n$ pieces $s_1,s_2,\dots,s_n$ over a finite field $\mathbb{F}_p$ such that $s=\sum_{i=1}^n s_n \bmod{p}$. Such additive secret sharing is the basic structure of many generic MPC protocols such as SPDZ \cite{damgaard2012multiparty}. Unlike additive secret sharing, Shamir secret sharing leverages non-linear mapping to reconstruct the secret. Specifically, for a $(t,n)$ Shamir scheme, to share a secret $s$, one needs to randomly select $t-1$ elements $a_1,a_2,\dots,a_{t-1}$ from a finite field $\mathbb{F}_p$ and let $a_0=s$, to construct the polynomial
$$f(x)=a_0+a_1x+a_2x^2+\dots+a_{t-1}x^{t-1} \bmod{p}$$
Then one can $n$ distinct points on the curve defined by the Lagrange polynomial except for the point $(0,s)$ and distribute them as shares to $n$ parties. To reconstruct the secret $s$, once one has collected at least $t$ Shamir shares $(x_i,y_i)$, the constant term of the above Lagrange polynomial can be obtained by calculating
$$s=f(0)=\sum_{j=0}^{t-1}y_j\prod_{m=0, m\neq j}^{t-1}\frac{x_m}{x_m-x_j}$$
We should note that although the Shamir scheme involves non-linear mapping to reconstruct the secret, operations on its shares still hold additive homomorphism. Besides, the threshold structure of Shamir schemes makes it natural to be used to handle dropped users and to construct verification protocols. 

\subsection{Differential Privacy}
\label{sec:dp}
Differential Privacy (DP) is a technique that gives a solution to the paradox of learning knowledge from a large dataset but securing the privacy of individual participants \cite{dwork2014algorithmic}. 
Referring descriptions in \cite{dwork2006our,dwork2006calibrating}, the definition of DP can be summarized as: A randomized mechanism $\mathcal{M}$ is differentially private if for any two neighboring databases (NB) $\mathcal{X}$ and $\mathcal{X'}$, and for all possible outputs $S \subseteq \mathbb{R}$, it satisfies $\epsilon$-DP when
\begin{equation}
\label{eq:DP}
    \frac{P[\mathcal{M}(\mathcal{X})\in S]}{P[\mathcal{M}(\mathcal{X'})\in S]}\leq \exp(\epsilon).
\end{equation}
Here, $\epsilon$ is the privacy budget that controls the difference degree of the two outputs from $\mathcal{M}$ with the two NB as inputs. The smaller the $\epsilon$ is, the higher privacy level of the participant get, but lower of the utility.
Databases are NB when they follow any of the two conditions:
(1) if $\mathcal{X}$ and $\mathcal{X'}$ are two datasets that have at most one record different;
(2) if $\mathcal{X}$ and $\mathcal{X'}$ have one entry different.

The function sensitivity $S_\mathcal{M}$ of a randomized mechanism $\mathcal{M}$ can be represented as
    $S_\mathcal{M}= \mathop{max}\limits_{\mathcal{X},\mathcal{X'}}\left\|\mathcal{M}(\mathcal{X})-\mathcal{M}(\mathcal{X'})\right\|_1$,
which measures the maximum difference of the outputs when input a pair of NB. Here we use the ${\ell}_1$ sensitivity, it can also be other distance calculation methods based on different setting requirements.

In FL, the information privacy in a pair of NB $\mathcal{X}$, $\mathcal{X'}$ can be protected by adding random noise to the data or different model parameters based on a particularly selected differentially private-mechanism $\mathcal{M}$ \cite{dwork2006calibrating,abadi2016deep}. Some common used noise generation mechanisms include Gaussian mechanism \cite{dong2019gaussian} and Laplace mechanism \cite{dwork2006calibrating}.

There are different definitions for the concepts of central/global DP and local DP. 
To make these two concepts clearer and easier to understand, we distinguish them in this paper from the perspective of who adds noise to the FL training parameters.
The approaches that perturb the parameters by the global server or a trusted third party server are regarded as GDP-based works and the techniques that add noise by local users are treated as LDP-based methods. GDP follows Eq.(\ref{eq:DP}). LDP is held by $
\frac{P[\mathcal{M}_n(\mathcal{X}_n)\in S_n]}{P[\mathcal{M}_n(\mathcal{X}'_n)\in S_n]}\leq \exp(\epsilon_n)$ for any $\mathcal{X}_n$ and $\mathcal{X}'_n$, where $n$ means the $n$-th participant.
The nature of differential privacy can promise individual-indistinguishable, which makes all types of DP-based techniques have the ability to prevent membership inference attacks. GDP can also bound the success of property inference, but it will lead to huge loss of the FL model utility if without a large number of local users \cite{melis2019exploiting,naseri2020local}. GDP-based PPAgg approach in FL can add less noise than the LDP-based PPAgg method when guaranteeing the same privacy-preserving level but it should satisfy the condition that the global server is trusted or a trusted third party server is available. 
LDP-based PPAgg is more common and practical in FL but this kind of approach can not restrict the property inference attacks.  
There are also some new DP settings emerging with practical privacy loss accounting method, like ($\epsilon$, $\delta$)-DP \cite{dwork2014algorithmic}, different versions of concentrated differential privacy (CDP \cite{dwork2016concentrated}, zCDP \cite{bun2016concentrated}, tCDP \cite{bun2018composable}), and R\'e{n}yi differential privacy (RDP) \cite{mironov2017renyi}. They are variants of the standard DP definition.
    
    

\subsection{Blockchain}
\label{sec:blockchain}

Blockchains was originated from the concept of cryptocurrencies, i.e., Bitcoin, to serve as a tamper-proof and decentralized ledger to record an ordered set of transactions in a transparent and immutable manner \cite{nakamoto2008bitcoin}. Specifically, these transactions are verified by trustless blockchain nodes through a decentralized consensus protocol and are constructed into blocks before attaching to the blockchain. Apart from the transactions, a block also contains a cryptographic hash of the previous block, which provides linkability and traceability. Here, we summarize the key advantages that blockchain networks can offer as follows:
\begin{itemize}
    \item \textit{Decentralization}: Each transaction to be attached to the blockchain must be confirmed upon the agreement among the majority of the blockchain nodes through a decentralized consensus protocol. As such, the single-point-monopoly of a centralized network can be removed from the blockchain.
    \item \textit{Immutability}: The transactions stored in blockchain ledgers cannot be altered or tampered with unless the majority of nodes are compromised. Such security is guaranteed by the cryptographic techniques used in the blockchain that any change of the transaction data can be observed by all blockchain nodes.
    \item \textit{Transparency} and \textit{Auditability}: The transactions stored in blockchain ledgers are visible to all blockchain nodes and can be traced back for verification.
    \item \textit{Pseudonymity}: By using the digital signature techniques, blockchain allows nodes to execute the transaction in an anonymous manner, without intervention of any trusted third-party.
\end{itemize}

To enable more complicated function evaluation, e.g., aggreation, over a blockchain rather than only data recording, smart contract \cite{wood2014ethereum} is utilized which can be written into lines of code and automatically executed when pre-defined conditions are met. In general, the smart contract cannot be modified once it is deployed in the blockchain, and its execution is also decentralized, which ensures stable and reliable control functions \cite{wu2020fedbc}.

\subsection{Trusted Execution Environment}
\label{sec:tee}
The Trusted Execution Environment (TEE), as defined by GlobalPlatform, is a secure area of the main processor that allows the sensitive data and code to be stored, processed, and protected in an isolated and trusted environment \cite{platform2018introduction}. In other words, TEE is isolated from the pure software environment, i.e., Rich Operating System Execution Environment (REE). Thus, TEE guarantees the confidentiality and integrity of insider applications and related data against any attacks from REE. 

The implementations of TEE, e.g., \cite{kostiainen2012board, optee}, are supported by hardware enclaves, such as Intel SGX \cite{mckeen2013innovative} and ARM TrustZone \cite{arm2009arm}, where a trade-off exists between the computation resource, e.g., limited memory size, and provided security level. Note that compared to REE-based applications, TEE usually involves extra costs regarding hardware which may hinder its large-scale deployment.



\section{Privacy-Preserving Aggregation Protocols in Federated Learning}
\label{sec:ppa-protocols}
In this section, we will survey different constructions of PPAgg protocols, their applications in FL systems, and an extensive analysis of the advantages and disadvantages of these selected PPAgg protocols and solutions.

\subsection{Masking-based Aggregation}
\label{sec:masking-based aggregation}

\textbf{Pair-wise masking.}
As mentioned in Section \ref{sec:one-time pad}, OTP-based masking techniques can be adopted to encrypt a message to fully preserve its privacy. For example, in a typical federated learning system, the users can mask their models and then upload them to the central server for aggregation. This requires well-designed protocols to enable the central server to obtain the aggregation results from these masked models. This research direction is arguably pioneered by the design of SecAgg, where the authors propose a pair-wise additive masking technique that the masks can be automatically canceled when the FL users' masked models are aggregated. In specific, assume there is set of ordered FL users $\mathcal{U}$ where each $u_i \in \mathcal{U}$ has a vector $\boldsymbol{x}_i$. In SecAgg protocol, each user $u_i$ add a pair-wise additive mask to its held vector $\boldsymbol{x}_i$ to get the masked vector $\boldsymbol{y}_{i}$.
$$\boldsymbol{y}_{i}=\boldsymbol{x}_{i}+\sum_{i<j} \text{PRG}(s_{i, j})-\sum_{i>j} \text{PRG}(s_{j, i})$$
where pseudorandom generator (PRG) can randomly generate a sequence numbers based on the seed $s_{i, j}$. It is obvious from the above equation that the masks will be canceled when all masked vectors $\boldsymbol{y}_i$ are added such that 
\begin{footnotesize}
\begin{equation*}
\begin{aligned}
\sum_{u_i\in \mathcal{U}}\boldsymbol{y}_{i}=\sum_{u_i\in \mathcal{U}}\left(\boldsymbol{x}_{i}+\sum_{i<j} \text{PRG}(s_{i, j})-\sum_{i>j} \text{PRG}(s_{j, i})\right)=\sum_{u_i\in \mathcal{U}} \boldsymbol{x}_{i}    
\end{aligned}
\end{equation*}
\end{footnotesize}
In addition, to handle the dropped FL users during the protocol execution, the Shamir secret sharing scheme (see Section \ref{sec:mpc} and the related security constraints discussed in \cite{zhao2021information}) is used to secretly share the seeds among users. Diffie-Hellman (DH) key exchange protocol \cite{diffie2019new} is adopted to make an agreement on the seed $s_{i, j}$ for each pair of user $(u_i,u_j)$. Note that the process of seed agreement is necessary as with the help of seeds and PRG, each FL user only needs to share the seeds with others rather than the whole masked vector, which greatly reduces the communication overheads especially for handling dropped users. Experimental results in \cite{zheng2022aggregation} can support this point. For simplicity, here we omit the introduction of double-masking, consistency-check, and other techniques in SecAgg to improve the security, and we refer interested readers to \cite{bonawitz2017practical} for the details.

Note that the SecAgg scheme is not cost-effective for large-scale federated learning applications. For an $n$-user FL system, it requires $\mathcal{O}(n^2)$ communication-round to run the pair-wise DH key exchange protocol. Therefore, communication-reduction techniques from an ML perspective are introduced to combine with SecAgg to further reduce the overheads. For example, in \cite{zheng2022aggregation,kairouz2021distributed}, well-designed quantization techniques are used to optimize the communication efficiency. In \cite{bonawitz2019federated}, the authors integrate the random rotation technique with SecAgg to aggressively adjust the quantization range of the users' models to reduce the model volume. CodedPaddedFL \cite{schlegel2021codedpaddedfl} adopts coding technique to improve the efficiency. In \cite{elkordy2022heterosag}, heterogeneous quantization is introduced to adjust users' quantization level according to their available communication resources. In \cite{ergun2021sparsified}, gradient sparsification technique is adopted to compress the users' model. Besides, SecAgg-based PPAgg protocols for federated submodel learning can be found in \cite{cui2021practical} and \cite{niu2019secure}. However, the above-mentioned works rely on the SecAgg scheme for aggregation, and thus still involve high communication overheads when it comes to large-scale FL systems.

To reduce the communication overheads of SecAgg while keeping the use of the pair-wise masking technique, several follow-up schemes are proposed in which FL users communicate across only a subset of the user rather than all users. For example, TurboAgg \cite{so2021turbo} divides $n$ FL users into $n/\log n$ groups and then follows a multi-group circular structure for aggregation. In this case, each FL user in a group communicates with only the users in the next group. A similar grouping structure can be found in SwiftAgg \cite{jahani2022swiftagg}. However, these schemes require additional communication rounds to process between groups and sacrifice some security guarantees against colluding adversaries.

\begin{figure}[htbp]
    \centering
    \includegraphics[width=0.8\linewidth]{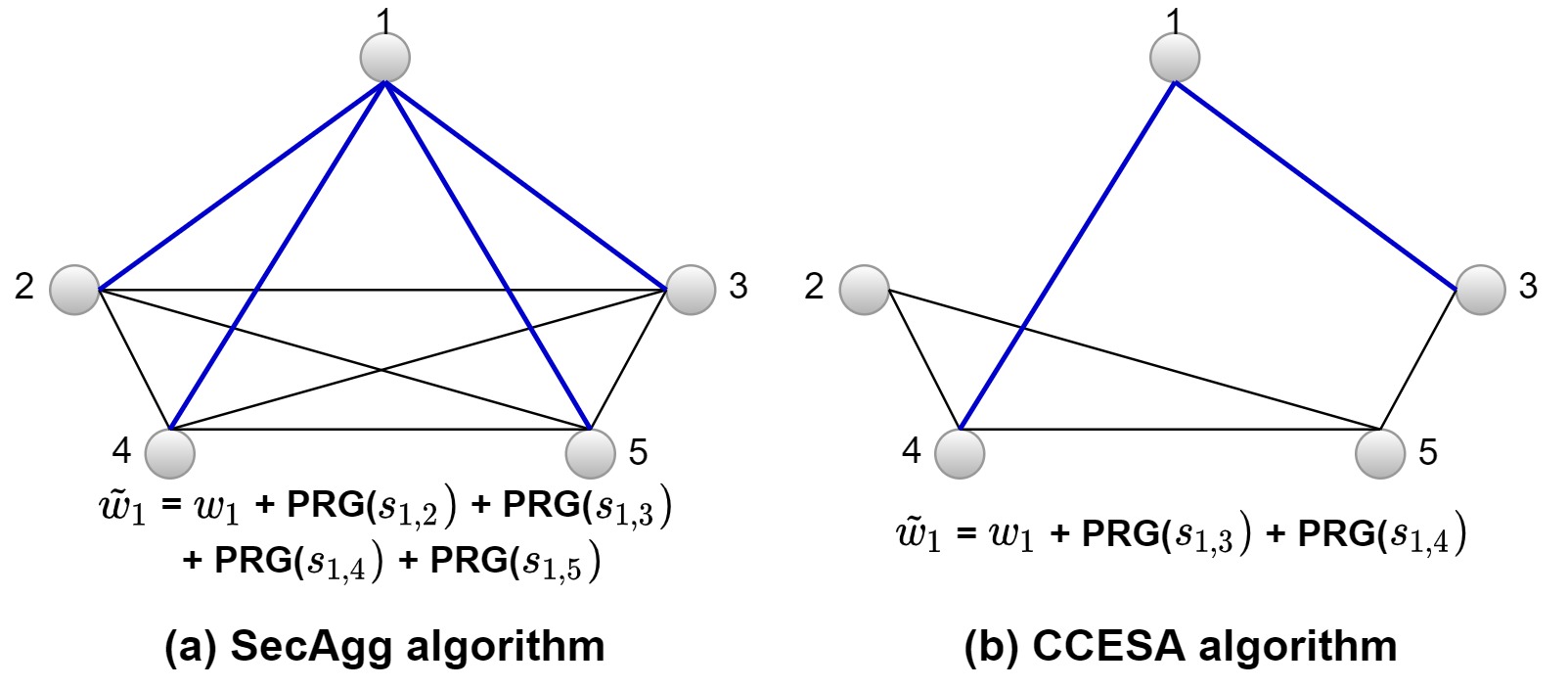}
    \caption{The sparse communication graph of CCESA \cite{choi2020communication} compared to the complete communication graph of SecAgg \cite{bonawitz2017practical}.}
    \label{fig:ccesa}
\end{figure}

As such, aggregation schemes with a non-group architecture are considered. In CCESA \cite{choi2020communication}, the authors demonstrate that the scheme in which FL users communicate, i.e., run the pair-wise DH key exchange protocol, over a sparse random graph instead of the complete graph provides a similar security guarantee to that of the SecAgg scheme but with a lower communication overhead. An illustrative topology comparison between CCESA and SecAgg is given in Figure \ref{fig:ccesa}. CCESA \cite{choi2020communication} implements the sparse random graph by a Erd{\"o}s-R{\'e}nyi graph. In such a graph, each pair of FL users is connected with a probability $p$. Therefore, the selection of $p$ leads to a trade-off between the security level and protocol efficiency. A proper $p$ given in \cite{choi2020communication} that provides similar security guarantee to that of SecAgg \cite{bonawitz2017practical} reduce the communication complexity from $\mathcal{O}(n^2)$ to $\mathcal{O}(n\log n)$. Another independent work SecAgg+ \cite{bell2020secure} adopts the Harray graph to replace the complete graph for the communication of the pair-wise masking. Different from Erd{\"o}s-R{\'e}nyi graph, Harray graph is a $k$-connected graph with graph vertices having the smallest possible number of edges. Similarly, the proper $k$ selected in SecAgg+ also leads to a $\mathcal{O}(n\log n)$ communication complexity.

Another direction to improve the efficiency of the pair-wise-masking-based aggregation scheme is to replace the DH key exchange protocol with a lightweight or non-interactive algorithm. For example, in Nike \cite{mandal2018nike}, with the assistance of two non-colluding cryptographic secret providers, i.e., aided servers, the seed of each pair of FL users can be generated in a non-interactive way using a bivariate polynomial. In the scheme proposed in \cite{ShiCRCS11} which can be adapted to the aggregation in FL, a trusted dealer is involved to assign seeds to users of which the sum equals zero. In FLASHE \cite{jiang2021flashe}, with the assumption that the semi-honest central does not collude with any single FL user, the authors propose a lightweight homomorphic encryption algorithm with a pair-wise masking style, which achieves much efficiency improvement. However, the trust distribution of Nike \cite{mandal2018nike} and FLASHE \cite{jiang2021flashe} are limited to the number of aided servers and non-colluding assumptions, hence limiting their application scenarios. Note that to deal with dropped users, i.e., cancel their masks, all the users' seeds are secretly shared, e.g., using Shamir's secret sharing scheme in SecAgg, such that a set of alive users can reconstruct the seeds of dropped users to cancel their masks. Thus, secret sharing results in both dropout resilience and higher overheads. Alternatively, a recent work \cite{zheng2022aggregation} improves the efficiency upon SecAgg by removing the operations for secretly sharing seeds between FL users, resulting in a much lower communication cost if there is no dropped user, but a higher communication cost for the case that involve a large number of dropped users. Similar to \cite{zheng2022aggregation}, the authors of \cite{wang2021feverless} remove seed secret sharing operations but require all alive FL users to upload the shares of the whole masking vector to handle dropped users.

\begin{figure}[htbp]
    \centering
    \includegraphics[width=0.8\linewidth]{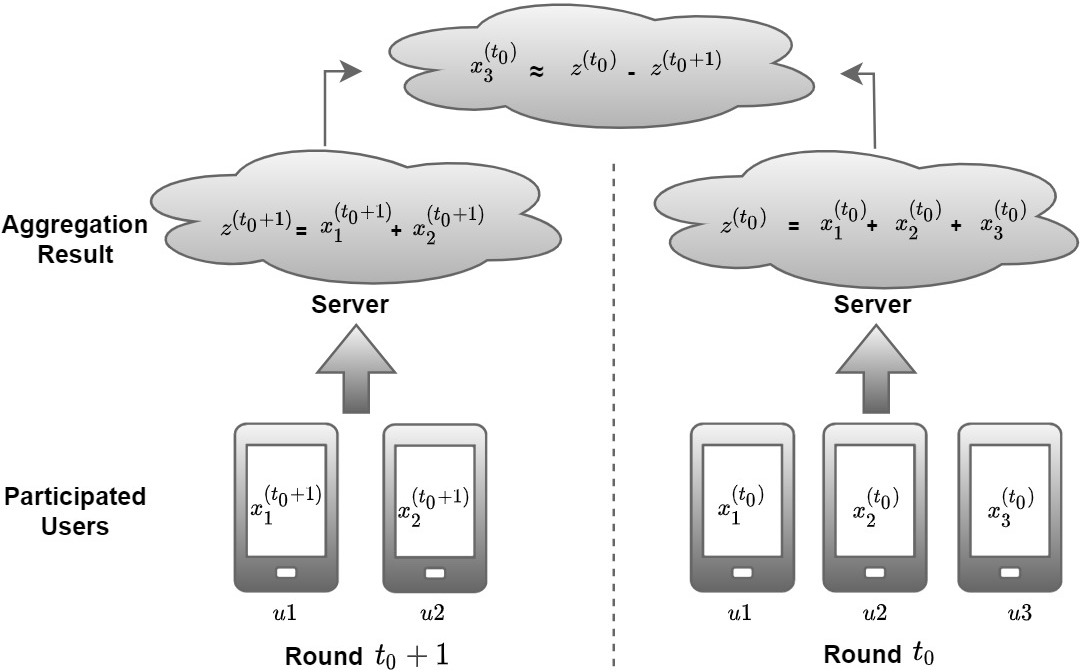}
    \caption{An adaptive figure from \cite{so2021securing} to describe the information leakage due to the multi-round FL training.}
    \label{fig:mrsa}
\end{figure}

Apart from protecting user privacy in a single FL round, several studies focus on the privacy issues caused by multiple-round FL training. For example, FedBuff \cite{nguyen2021federated} and LightSecAgg \cite{so2021lightsecagg} allow asynchronous aggregation of which the security can be enhanced by integrating existing PPAgg protocols. In a recent work \cite{so2021securing}, the authors point out that even with the aforementioned privacy-preserving aggregation protocols, the multiple-round FL training may lead to severe information leakages due to the dynamic user participation. As shown in Figure \ref{fig:mrsa}, user $u_1,u_2,u_3$ participate in round $t$, and user $u_1,u_2$ participate in round $t+1$. If the model of $u_1$ and $u_2$ do not change significantly over the two rounds, the server can reconstruct the model of $u_3$ with a very small error. This privacy issue should be noted as it is common in FL training when the global FL model converges. In \cite{so2021securing}, a naive mitigation method is proposed that requires the aggregation from a set of user batches rather than the individual users, hence adversaries cannot differentiate the user models in the same batch for any long time.

\begin{figure}[htbp]
\setlength{\abovecaptionskip}{-0.1cm} 
    \centering
    \includegraphics[width=0.65\linewidth]{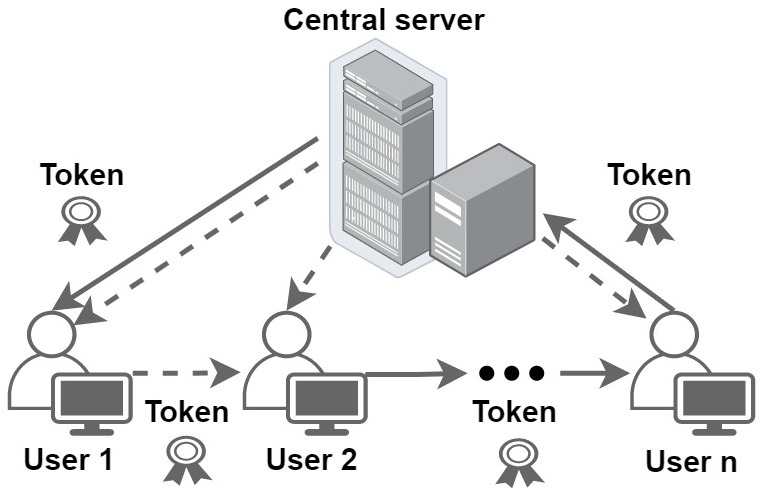}
    \caption{Aggregation following a chain structure.}
    \label{fig:one-shot chain}
\end{figure}

\textbf{Non-pair-wise masking.}
Although the pair-wise masking technique provides the attractive property that masks can be canceled when aggregating all the masked vectors, it naturally involves the interactions for the seed agreement between pairs of FL users, and thus hinders efficient deployments for large-scale FL systems. Therefore, some recent works replace pair-wise masking with lightweight non-pair-wise masking followed by one-shot unmasking. More specifically, in such a scheme with $n$ users, each FL user $u_i$ generates its mask $r_i$ without any interaction with others and uses it to encrypt its held vector $x$, then uploads the masked vector, e.g., $y_i=x_i+r_i$ in an additive masking manner, to the central server. Meanwhile, all users involve a protocol that allows the central server to know the sum of users' masks $R=\sum r_i$ while keeping the privacy of each $r_i$. In this case, the server is able to obtain the aggregation result by calculating $\sum x_i=\sum y_i-R$. For example, in HyFed \cite{nasirigerdeh2021hyfed}, a trusted party is involved to calculate the aggregated noise from users, which then be sent to the server for one-shot unmasking. In \cite{liu2022efficient}, homomorphic PRG (HPRG) is adopted to achieve a lightweight mask generation, hence greatly reducing the communication overheads. Besides, relying on the multiplicative masking and the additive homomorphic property of Shamir secret sharing scheme and HPRG, the server can obtain the seed for canceling the masks without knowing any single user's seed, hence canceling the mask to obtain the correct aggregation result. Similar HPRG-based scheme is proposed in \cite{liu2021efficient}.  The idea of one-shot unmasking is also employed in \cite{zhao2021information} where a trusted third party coordinates dropped users and assists to calculate for unmasking. In the follow-up work LightSecAgg \cite{so2021lightsecagg}, the authors propose a lightweight and dropout-resilience secret sharing method, hence removing the trusted third party in \cite{zhao2021information} and allowing the server to do one-shot unmasking based on the received shares from alive FL users. Some other schemes achieve one-shot unmasking with a chain structure \cite{li2020privacy,ge2021chain,chen2021ppt,sandholm2021safe}. In specific, as shown in Figure \ref{fig:one-shot chain}, assume a total order on all FL users $u_i$, a central server or leader generates a mask and assigns it to the first user $u_1$, which is used by $u_1$ to mask its model. After that, $u_1$ transfer its masked model to the user $u_2$, then $u_2$ adds its model to the masked model received from $u_1$ and transfer it to $u_3$. Following this way, all users aggregate their models one by one in a chain style and the last user transfers its result to the central server for unmasking. However, these works assume that the central server is trusted and there is no collusion between the server and users, which is not practical for real-life FL applications.

\begin{table*}
\scriptsize
\centering
\caption{A summary of masking-based PPAgg protocols.}
\label{tab:masking-agg}
\begin{tabular}{|c|c|c|c|c|c|} 
\hline
\textbf{Scheme} 
& \begin{tabular}[c]{@{}c@{}}\textbf{Masking }\\\textbf{type}\end{tabular}         & \textbf{Threat model}        
& \begin{tabular}[c]{@{}c@{}}\textbf{Privacy}\\\textbf{guarantee}\end{tabular}                             
& \textbf{Complexity discussion}      
& \begin{tabular}[c]{@{}c@{}}\textbf{Methods for security and} \\\textbf{efficiency improvements}\end{tabular}  \\ 
\hline
\cite{bonawitz2016practical} \cite{bonawitz2017practical}   
& \multirow{26}{*}{\begin{tabular}[c]{@{}c@{}}Pair-wise\\mask\end{tabular}}    
& \multirow{13}{*}{\begin{tabular}[c]{@{}c@{}}Malicious users \\and server~\end{tabular}} 
& \multirow{20}{*}{Local model}               
&\multirow{5}{*}{\begin{tabular}[c]{@{}c@{}}Impractical for large-scale \\FL due to~$\mathcal{O}(n^2)$\\communication complexity\end{tabular}}
& Baseline                          
\\ 
\cline{1-1}\cline{6-6}
\cite{zheng2022aggregation}
&&&&
& Model Quantization;TEE\\ 
\cline{1-1}\cline{6-6}
\cite{kairouz2021distributed} \cite{bonawitz2019federated} \cite{elkordy2022heterosag}
& &&&                                                                         
& Model Quantization
\\ 
\cline{1-1}\cline{6-6}
\cite{ergun2021sparsified}       
&&&&                                                                         
& Coding approach                                                                          
\\ 
\cline{1-1}\cline{6-6}
\cite{schlegel2021codedpaddedfl}                                                   
&&&&
&Model spasification                                                                      
\\ 
\cline{1-1}\cline{5-6}
\cite{cui2021practical} \cite{niu2019secure}                     
&&&                                             
& Determined by the size of submodels                                                                        
& Submodel aggregation 
\\ 
\cline{1-1}\cline{5-6}
\cite{so2021turbo}~\cite{jahani2022swiftagg} &&&
&\begin{tabular}[c]{@{}c@{}}Requires additional $\mathcal{O}(n/\ \log n)$\\communication rounds\end{tabular}                                                                        & Group aggregation                                                                             \\ 
\cline{1-1}\cline{5-6}
\cite{choi2020communication}~\cite{bell2020secure}              &&&                                             
&\begin{tabular}[c]{@{}c@{}}Reduce communication\\complexity to $\mathcal{O}(\log n)$\end{tabular}         
& Sparse communication graph   
\\ 
\cline{1-1}\cline{5-6}
\cite{mandal2018nike}~\cite{ShiCRCS11}                          &&&                                             
&\begin{tabular}[c]{@{}c@{}}Replace user interactions\\with a trusted party\end{tabular}       
& Trusted third party  \\ 
\cline{1-1}\cline{5-6}
\cite{zheng2022aggregation}&&&                              
& Efficient with few dropped users       
&\begin{tabular}[c]{@{}c@{}}Removing secret sharing\\and introducing TEE\end{tabular} \\ 
\cline{1-1}\cline{3-3}\cline{5-6}
\cite{wang2021feverless}&                                     
&\multirow{3}{*}{\begin{tabular}[c]{@{}c@{}}Semi-honest users\\and server\end{tabular}}
&& Efficient with low dropout rates
&\begin{tabular}[c]{@{}c@{}}Replacing seed secret\\sharing with sending vectors\end{tabular}      
\\ 
\cline{1-1}\cline{5-6}
\cite{nguyen2021federated}~\cite{so2021lightsecagg}
&&&
&\begin{tabular}[c]{@{}c@{}}Determined by the\\integrated PPAgg protocol\end{tabular}        
&\begin{tabular}[c]{@{}c@{}}Asynchronous aggregation and\\lightweight reconstruction\end{tabular}   
\\ 
\cline{1-1}\cline{3-3}\cline{5-6}
\cite{jiang2021flashe}                                                                                                                                                                                                          &                                     & \begin{tabular}[c]{@{}c@{}}Semi-honest server\\~and users\end{tabular}            &                                             &  Efficient in M1 setting                                                                       & Lightweight HE                                                                  \\ 
\cline{1-1}\cline{3-6}
\cite{so2021securing}&               
&\multirow{3}{*}{\begin{tabular}[c]{@{}c@{}}Semi-honest users\\ and server\end{tabular}}                        
& \begin{tabular}[c]{@{}c@{}}Multi-round privacy\end{tabular}
& Determined by the user selection strategy        
& Batch partitioning          
\\ 
\cline{1-1}\cline{4-6}
\cite{feng2020practical}&                                     &                                                                                         & Global model                               & Efficient without protecting local models                                                                        & Mask global model                                                                        \\ 
\cline{1-1}\cline{4-6}
\cite{mandal2019privfl}                                                                                                                                                                                                        &                                     &                                                                                         & \multirow{3}{*}{\begin{tabular}[c]{@{}c@{}}Local model;\\Global model\end{tabular}} & Determined by the 2PC protocol adopted                                                                        & Two-party computation                                                              \\ 
\cline{1-1}\cline{3-3}\cline{5-6}
\cite{mugunthan2019smpai}&                                     
& \begin{tabular}[c]{@{}c@{}}Semi-honest users\\and server\end{tabular}                                                                                    &                                             & Determined by the MPC protocol adopted                                                                         & MPC and central DP                                                                                      \\ 
\cline{1-1}\cline{3-6}
\begin{tabular}[c]{@{}c@{}}\cite{xu2019verifynet} \cite{han2021verifiable}\\ \cite{guo2020v}\end{tabular}&
&\begin{tabular}[c]{@{}c@{}}Semi-honest users \\and server\end{tabular}
&\multirow{5}{*}{Local model}
&\multirow{5}{*}{\begin{tabular}[c]{@{}c@{}}Involve additional complexity\\ for verification based\\ the selected techniques.\end{tabular}}
&\multirow{5}{*}{\begin{tabular}[c]{@{}c@{}}Methods of guarantee\\the aggregation correctness\end{tabular}}\\
\cline{1-1}\cline{3-3}
\cite{hahn2021versa}
&&\begin{tabular}[c]{@{}c@{}}Semi-honest users;\\Malicious server\end{tabular}
&&&\\
\cline{1-1}\cline{3-3}
\cite{burkhalter2021rofl} 
&&\begin{tabular}[c]{@{}c@{}}Malicious users;\\Semi-honest server\end{tabular}
&&&\\
\hline
\cite{liu2022efficient} \cite{liu2021efficient}& \multirow{5}{*}{\begin{tabular}[c]{@{}c@{}}Non\\pair-wise\\mask\end{tabular}} 
&\begin{tabular}[c]{@{}c@{}}Malicious users \\and server\end{tabular}
& \multirow{5}{*}{Local model}                                 & Efficient for small ML models                                                                        & Homomorphic PRG                                                                                     \\ 
\cline{1-1}\cline{3-3}\cline{5-6}
\cite{nasirigerdeh2021hyfed} \cite{zhao2021information}&        
& \multirow{3}{*}{\begin{tabular}[c]{@{}c@{}}Semi-honest users \\and server\end{tabular}}                                                                                        &                                             &\begin{tabular}[c]{@{}c@{}}Replace complicated unmasking\\with a trusted party\end{tabular}                                                                         & Trusted third party                                                                      
\\ 
\cline{1-1}\cline{5-6}
\begin{tabular}[c]{@{}c@{}}\cite{li2020privacy}  \cite{ge2021chain}\\ \cite{chen2021ppt} \cite{sandholm2021safe}\end{tabular}                                                   &                                     &                                                                                         &                                             &  \begin{tabular}[c]{@{}c@{}}Additional $n$ communication \\rounds for following the chain structure \end{tabular}                                                                        & Chain structure                                                                          
\\
\hline
\end{tabular}
\end{table*}

\textbf{Protecting the global model.}
Note that the above-mentioned masking-based aggregation protocols aim to protect the privacy of FL users' models or gradients, hence their raw data. However, to protect the privacy of the global FL model, additional privacy-preserving techniques need to be integrated. In \cite{feng2020practical}, the server masks the global model and requires all FL users to do the training based on the masked model. After the training, each user sends some supplementary information to the server for unmasking. Note that such a scheme does not provide additional privacy protection on users' model compared to the standard FL scheme. In PrivFL \cite{mandal2019privfl}, a two-party computation technique is adopted, which allows each server-user pair to jointly train an ML model while preserving the privacy of both the global model and user's local model. After that, each user masks its share and sends it to the server, and all users are involved in a privacy-preserving protocol, e.g., SecAgg \cite{bonawitz2017practical} or SecAgg+ \cite{bell2020secure}, to aggregate the sum of their masks, which then be used by the server for unmasking. Furthermore, some other techniques are considered for integration to improve the security during the whole aggregation. For example, in \cite{mugunthan2019smpai}, differentially private noises are added to users' models to guarantee privacy during weighted averaging in aggregation. TEE in \cite{zheng2022aggregation}, pseudo-random functions in \cite{xu2019verifynet}, MAC-like technique in \cite{han2021verifiable}, homomorphic hash in \cite{hahn2021versa}, zero-knowledge proofs in \cite{burkhalter2021rofl}, and commitment scheme in \cite{guo2020v} are deployed to guarantee that the server correctly aggregates the sum from FL users.

So far, we have reviewed masking-based aggregation protocols to protect users' model privacy and global model privacy. We should note that aggregation protocols based on pair-wise masking allow efficient unmasking with dropped users, and thus are suitable for the cross-device setting where FL users are mobile IoT devices that may drop out of the system at any time. In contrast, aggregation protocols based on one-shot unmasking usually result in better efficiency for cross-silo settings. In most cases, there is a trade-off between security assumptions, e.g., a trusted party or non-colluding participants, and efficiency, e.g., computation, communication, or storage costs. Besides, to provide additional security apart from a privacy perspective, other cryptographic tools and trusted environments need to be considered. To summarize, we list the aforementioned aggregation schemes with related features in Table \ref{tab:masking-agg}.

\subsection{HE-based Aggregation}
\label{sec:he-based aggregation}
\begin{figure}[htbp]
    \centering
    \includegraphics[width=0.65 \linewidth]{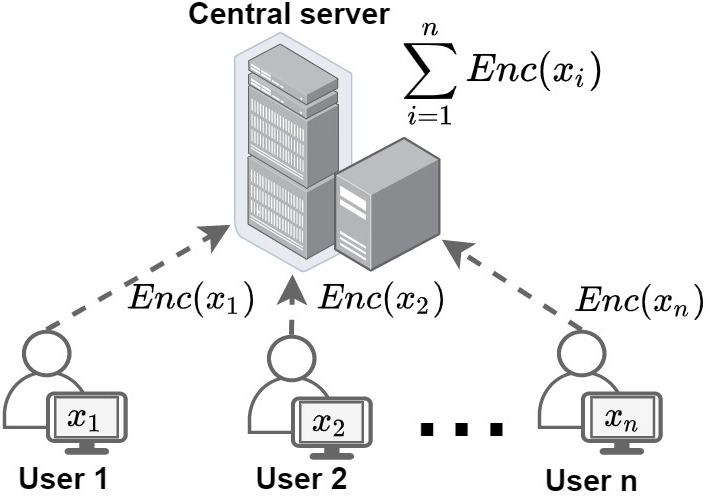}
    \caption{An illustrative figure for the aggregation in FL based on homomorphic encryption.}
    \label{fig:he-agg}
\end{figure}

The HE-based aggregation in FL is quite straightforward than that of masking-based aggregation. In general, to aggregate the sum of users' locally trained models in an FL round, the users just need to encrypt their models and send them to the central server. Then the central server adds the received encrypted models together relying on the additive homomorphic property of the used crypto-system, which can be decrypted to obtain the global model in that FL round (see Figure \ref{fig:he-agg}).

The security of HE-based aggregation protocols in FL is achieved through their underlying crypto-system. According to the introduction of homomorphic encryption systems in Section \ref{sec:he}, the homomorphic property is held only when the ciphertexts are encrypted using the same public key. Therefore, in a distributed setting such as FL, the key management of the crypto-system usually determines the threat model and application scenarios. More specifically, for a public-key crypto-system used for aggregation in FL, the ownership of the secret key is the most important factor to be considered when it comes to industrial deployments. First, we summarize the settings of the secret key management in FL.

\begin{itemize}
    \item \textbf{M1}: Only known to all FL users and not to the central server.
    \item \textbf{M2}: Only known to the central server and not to any FL users.
    \item \textbf{M3}: Split across all or a set of FL participants.
\end{itemize}

\textbf{M1 setting.} In the M1 setting, the secret key is known to all FL users but kept confidential against the central server. In this case, users' model privacy is protected but the global model is public to all FL participants. The secret key can be generated via interactions between users \cite{aono2017privacy,dong2020eastfly,chen2021network} or with the assistance of a trusted third party \cite{zhou2020privacy,zhang2020privacy,fang2021privacyhomo,xu2021fedv,tang2019privacy}. Besides, the crypto-system can be instantiated in different way to provide different security level, e.g., whether post-quantum or not, such as RSA-based \cite{yang2020privacy}, BGN-based \cite{xu2021achieving}, ElGamal-based \cite{fang2021privacy}, Paillier-based \cite{ZhangLX00020,aono2017privacy,asad2020fedopt,guo2021privacy,dong2020eastfly,zhang2020privacy,xu2021fedv,zhang2021dubhe,guo2021privacy}, lattice-based crypto-system \cite{alexandru2020private,stripelis2021secure}. However, the above-mentioned schemes require rigorous security assumption that all users are at least semi-honest while there is no collusion between any user and the central server. This assumption is quite weak since the server can directly break the security of users' model privacy by creating a pseudonymous FL user.

\begin{figure}[htbp]
    \centering
    \includegraphics[width=0.7 \linewidth]{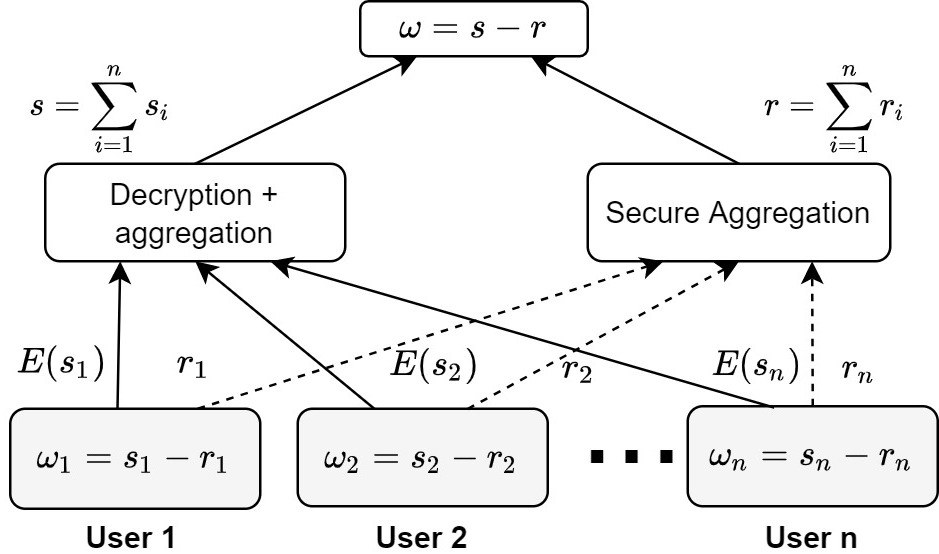}
    \caption{An adaptive figure from PrivFL \cite{mandal2019privfl} for illustration.}
    \label{fig:privfl}
\end{figure}

\textbf{M2 setting.} Different from the M1 setting that provides users' model privacy, in the M2 setting, the secret key is held by the central server, which aims to protect the privacy of the global model from the server. Such consideration is common in many FL applications as the central server is usually a commercial company or consulting agency that intends to use FL to improve its ML model performance and provides model inference as a service for commercial profits \cite{cheng2020federated,jin2020towards,yang2021toward}. In this case, the privacy of the global model is considered to be protected. However, as the central server holds the secret key, the encrypted model sent from FL users to the server can be decrypted by the server, which directly breaks the security of users' model privacy. Therefore, other privacy-preserving techniques are required to further improve the privacy guarantee. For example, in PrivFL \cite{mandal2019privfl} where only the server holds the secret key of the HE scheme, users mask their models before aggregation and involve in a SecAgg protocol to aggregate the masks for unmasking (see Figure \ref{fig:privfl}), hence protects the privacy of both users' models and the global model. Other works such as \cite{hardy2017private, gao2019privacy,zhou2020privacy,liu2020secure,yang2020privacy,zhang2019pefl,asad2021ceep} rely on a trusted party to manage the secret key. In PIVODL \cite{zhu2021pivodl}, an FL user is randomly selected as the key generator in each round, hence adding randomness to improve the security. However, the trust distributions of these works are still limited to only one party. Besides, since the global model is encrypted, FL users have to train their local model over ciphertext, hence involving FHE or MPC which may lead to impractical computation or communication overhead, especially in the cross-device setting.

\begin{figure}[htb]
\setlength{\abovecaptionskip}{1pt} 
    \centering
    \includegraphics[width=1 \linewidth]{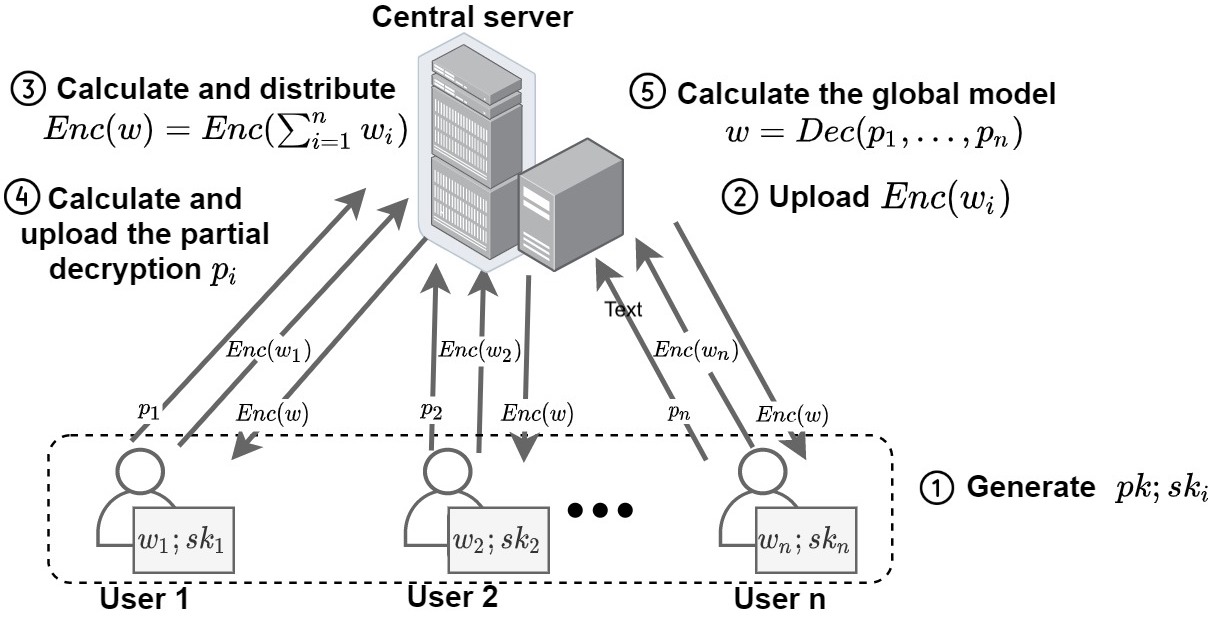}
    \caption{An illustrative workflow of the aggregation in FL with a threshold addtive HE crypto-system.}
    \label{fig:phe-fl}
\end{figure}

\textbf{M3 setting.} Recall that in the M1 setting, all or a set of users hold the same secret key, and thus the ciphertext is no longer meaningful as long as the aggregation server colludes with any user that has the secret key. A straightforward method to deal with this security issue is to split the secret key across a set of users, i.e., the M3 setting. In this case, with a threshold crypto-system (see Section \ref{sec:he}), several users that more than a threshold number must cooperate in order to decrypt an encrypted message. Such a setting improves the security guarantee upon that of the M1 setting as the central server must collude with much more FL users to break the security. For example, as shown in Figure \ref{fig:phe-fl}, in an FL system with a full-threshold crypto-system, all participants work as follows to achieve a privacy-preserving aggregation. (i) Users involve in a key generation protocol to generate a key-pair $(p_k,s_k)$ that the public key $p_k$ is known to everyone while the secret key $s_k$ is shared. For example, for an $n$-user FL system, each user $u_i$ has a partial secret key $s_k^i$ such that $f(s_k^1,\dots,s_k^n)=s_k$ where $f(\cdot)$ is determined according to the crypto-system used. (ii) Each user trains its local model $w_i$ based on the global FL model, and encrypts it using the public key $p_k$ and sends $Enc(w_i,p_k)$ to the central server. (iii) After received the encrypted models from users, the central server calculates $Enc(w,p_k)=\sum_{i=1}^n Enc(w_i,p_k)=Enc(\sum_{i=1}^n w_i,p_k)$ relying on homomorphic operations, and sends the result to users. (iv) Each user $u_i$ uses its partial secret key to calculate a partial decryption $p_i=Dec_p(s_k^i,Enc(w,p_k))$ and sends it to the server, where $Dec_p$ is determined according to the crypto-system used. (v) After received the partial decryptions from users, the server calculates $w=Dec(p_1,\dots,p_n)$ to obtain the aggregation result $w$, i.e., the global model of the current FL round, where $Dec(\cdot)$ is determined according to the crypto-system used.


Partial HE schemes are widely used for the M3 setting. For example, threshold Paillier crypto-systems are adopted in \cite{truex2019hybrid,liu2020boosting,li2021efficient,ma2022privacy}. In \cite{jiang2021secure}, a more efficient variant of threshold Paillier crypto-system called BCP is introduced. Upon basic threshold Paillier crypto-systems, Helen \cite{zheng2019helen} integrates zero-knowledge proof and some supporting protocols of SPDZ \cite{damgaard2012multiparty} to guarantee the security against active malicious FL participants. An ElGamal-based crypto-system is deployed in \cite{zhu2021distributed} in order to improve the computation efficiency compared to the Paillier-based crypto-system. A lightweight AHE scheme for aggregation in FL is proposed in \cite{tian2021secure}. However, the above-mentioned partial HE schemes support only addition or multiplication. Therefore, they are suitable to be used only for the aggregation part which involves only addition operations, instead of the functions that involve complex arithmetic operations.

\textbf{Protecting the global model.} To further provide privacy guarantees of the global FL model rather than only the users' models during aggregation, the global FL models are required to be encrypted. In this case, users have to train their local models over ciphertext. Since the training process involves both addition and multiplication operations, fully homomorphic encryption crypto-systems that support any arithmetic operations become necessary. Therefore, lattice-based FHE crypto-systems are adopted in FL schemes \cite{tian2021secure,hosseini2021secure,fereidooni2021safelearn,froelicher2021scalable,SavPTFBSH21} to enable more complicated function evaluation, e.g., ML model training, and extended to their multi-key versions for privacy purposes. For example, SAFELearn \cite{fereidooni2021safelearn} describes a general PPFL scheme based on a multi-key FHE crypto-system. SPINDLE \cite{froelicher2021scalable} proposes a PPFL scheme for the generalized linear ML model that protects the privacy of the whole FL workflow, i.e., the privacy of both users' models and the global model, relying on a multi-key version of CKKS crypto-system \cite{mouchet2020multiparty}. POSEIDON \cite{SavPTFBSH21} extends the supported ML models of SPINDLE \cite{froelicher2021scalable} from linear models to neural networks, and comes up with a distributed bootstrapping protocol for training deep neural networks in an FL setting. Taking into account that FL systems with a standard multi-key lattice-based FHE crypto-system do not allow new FL users to join who do not participate in the public key generation, \cite{hosseini2021secure} involves a setup phase based on Shamir secret sharing where all users can exchange their shares of secret keys, hence new users are allowed to join by obtaining corresponding shares. To improve the efficiency, some other works \cite{xu2019hybridalpha,wu2020towards,XuB00JL21} adopt multi-input functional encryption schemes. However, functional encryption involves high computation complexity for complicated functions and a trusted party is needed for key generation and distribution, which weakens the security guarantee compared to the threshold and multi-key crypto-systems based FL.

We should note that the nature of threshold and multi-key crypto-systems inevitably leads to an expensive public key generation process that involves interactions between all users. Besides, protecting the privacy of the whole FL workflow requires the users to train their ML models over ciphertext, which is impractical for complicated functions such as deep neural networks even with state-of-the-art techniques. Therefore, one needs to carefully weigh the trade-off between privacy guarantee and FL efficiency of such a setting according to the application scenario and its security requirements.


\subsection{MPC-based Aggregation}
\label{sec:mpc-agg}
\begin{figure}[htbp]
    \centering
    \setlength{\abovecaptionskip}{1pt} 
    \subfigure[Aggregation]{
    \begin{minipage}[b]{0.455\linewidth}
    \centering
    \includegraphics[width=\linewidth]{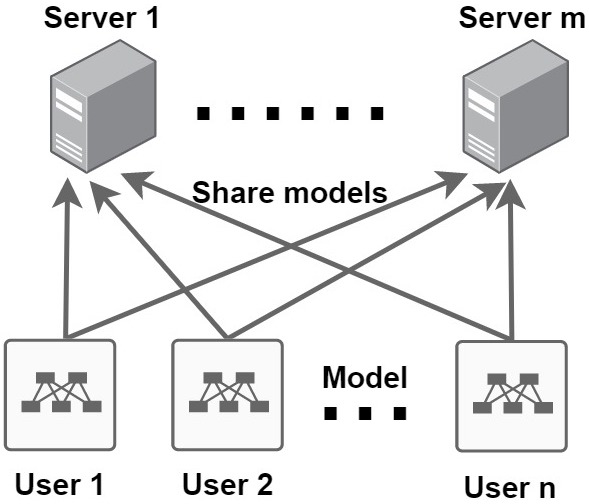}
    \label{fig:mpc-agg-agg}
    \end{minipage}}
    \subfigure[Training]{
    \begin{minipage}[b]{0.44\linewidth}
    \centering
    \includegraphics[width=\linewidth]{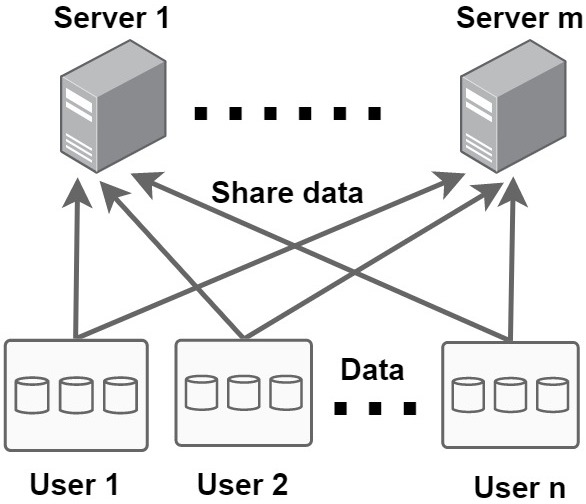}
    \label{fig:mpc-agg-trn}
    \end{minipage}}
    \caption{MPC-based privacy-preserving aggregation and training.}
    \label{fig:mpc-agg}
\end{figure}

\textbf{Share model.} For multi-party settings such as distributed machine learning and federated learning, MPC can be a natural option to enhance the security guarantee of the systems. A number of works employ MPC methods to achieve privacy-preserving aggregation and further privacy-preserving FL training. As shown in Figure \ref{fig:mpc-agg-agg}, MPC-based privacy-preserving aggregation protocols allow FL users to distribute, i.e., share (see Section \ref{sec:mpc}), their locally trained models to a set of agents, e.g., selected users or assistant servers. Then these agents jointly calculate the sum of users' models to obtain the share of aggregation result. After that, they may choose to reconstruct the result, i.e., the new global FL model. For example, in \cite{boer2020secure,sotthiwat2021partially}, each FL user shares its locally trained model to all users for aggregation using generic MPC protocols. In Fastsecagg \cite{kadhe2020fastsecagg}, by sacrificing some security, the authors substitute the standard Shamir secret sharing with a more efficient FFT-based multi-secret sharing scheme. Alternatively, models can be shared between two servers as in \cite{xu2022non,he2020secure,xu2020privacy,jayaraman2018distributed}, or several servers such as \cite{chen2020achieving,brunetta2021non}. Some other works introduce a two-phase secret-sharing-based aggregation \cite{kanagavelu2020two,zhu2020privacy}. In the first phase, all users are involved in an MPC-enabled selection protocol to construct a committee. Then in the second phase, all users share their models to the users in the committee for aggregation relying on standard MPC protocols that are similar to the aforementioned works. Furthermore, to guarantee the correctness of the aggregation result with a malicious central server, verifiable secret sharing schemes are adopted in \cite{fu2020vfl,brunetta2021non}.


\textbf{Share data.} Similarly, if FL users distribute their local data to several servers for training instead of aggregation, as shown in Figure \ref{fig:mpc-agg-trn}, the scheme boils down to the MPC-based distributed ML training. Note that such a scheme directly supports privacy-preserving aggregation as the locally trained models are also shared among MPC participants, i.e., servers or selected users. In general, such privacy-preserving training can be achieved by splitting the trust among two \cite{demmler2015aby,mohassel2017secureml}, three \cite{mohassel2018aby3,wagh2019securenn}, four \cite{byali2020flash,chaudhari2019trident}, or any number of servers, respectively, relying on generic MPC protocols. For example, \cite{sharma2019secure} adopts ABY \cite{demmler2015aby} and SPDZ \cite{damgaard2012multiparty} to achieve a secure federated transfer learning. \cite{chen2021homomorphic} and \cite{zheng2021towards} rely on a 2PC protocol. \cite{fang2021large} deploys a gradient tree boosting model based on GMW protocol. Multi-party settings are considered in \cite{zheng2019helen,ryffel2018generic} where SPDZ protocols are applied. Note that these works are the applications of generic MPC protocols to FL, of which the efficiency and security improvements mainly come from their underlying MPC schemes. We refer interested readers to \cite{hastings2019sok} for the details of state-of-the-art MPC protocols.

However, the nature of MPC limits its application to FL. Firstly, sharing secrets among all users usually leads to a large communication overhead. Thus, for a practical pure-MPC-based FL system, the data or models of FL users have to be transferred to a small number of MPC participants. In this case, security can be guaranteed only when the MPC participants do not collude or the majority of them are trusted. Furthermore, the above-mentioned MPC-based privacy-preserving training schemes usually require all MPC participants to keep online during the execution of the whole MPC protocol, and possess sufficient bandwidth and computational power. Therefore, these MPC participants cannot be resource-constrained mobile or IoT devices that may drop out of the system at any time, which is common in a cross-device FL setting.

\textbf{Handle dropped users.} Apart from the above-mentioned works that models or data are directly shared among MPC participants to protect privacy, secret sharing schemes can also be adopted as supporting protocols to handle dropped users or to verify computation results in FL systems. As described in Section \ref{sec:mpc}, a threshold secret sharing scheme, e.g., Shamir secret sharing, can be a natural choice to enable protocol execution even with a set of users of which the number is greater than a threshold value. If these users are set to be online users, the proposed protocol will obtain a dropout resilience. For example, in \cite{bonawitz2017practical,bell2020secure,liu2022efficient,choi2020communication}, the seeds for mask generation are shared using the Shamir scheme which allows the central server to reconstruct the masks of dropped users if the number of online users is greater than the threshold. If these users are set to be honest users, the proposed protocol will allow the verification of the computation results. For example, \cite{fu2020vfl,brunetta2021non} acknowledge the verification on the correctness of the aggregation result only when a set of users of which the number is greater than the threshold have verified the result. However, we should note that the setting of threshold value leads to a trade-off depending on its usage. A larger threshold means a stronger security guarantee but less protocol efficiency, and vice versa.

\subsection{DP-based Aggregation}
\label{sec:dp-based-aggregation}
Different from the data encryption approaches, DP-based aggregation is a data perturbation method that requires less computational overhead to achieve PPAgg in FL.
Data perturbation can be realized by adding noise to the FL training parameters according to different statistical data distribution mechanisms.
As introduced in \ref{sec:dp}, DP can be divided into GDP and LDP in FL according to who (the trusted curator/server or the data owner) operates the data perturbation process.
Both GPD and LDP PPAgg protocols have their own advantages and limitations.
In this subsection, we will review related research works that leverage DP-based techniques to secure information privacy in FL. 


\textbf{LDP.} Considering the practical situations in FL, it is more reasonable to assume that no trusted server is available. Hence, most previous research works protect the privacy of local users with LDP PPAgg protocols.
Local users perturb their information before sending the local updates to the global server.
In \cite{lu2019differentially}, the authors leverage the Gaussian mechanism to guarantee $\epsilon$ deferentially private for the local models to achieve secure and accurate resource sharing in the Internet of Vehicles (IoV) scene. 
Another work \cite{zhao2020local} also considers the privacy-preserving problem in IoV. The authors integrate their proposed novel LDP models with the FedSGD algorithm \cite{chen2016revisiting} to perturb the uploaded local gradients for securing the privacy information of the vehicle users. Their proposed LDP-FedSGD model can bound the success of membership inference and also realize better accuracy performance than the three compared models.
SFSL \cite{niu2020billion} applies LDP on submodel update to achieve plausible deniability for different mobile clients in an e-commerce recommendation FL system.
LDP-based PPAgg protocol with Gaussian noise in \cite{shi2021hfl} is applied both on the clipping client models and edge servers in a client-edge-cloud Hierarchical Federated Learning system.
Except for Gaussian noise based LDP-FL models mentioned above, there are also some research works that perturb the non-discrete parameters in local models with Laplace noise to prevent information leakage \cite{zhang2019efficient,wang2021protecting}.
Since most LDP-based PPAgg protocols need to add noise to the target parameters in each iteration for FL training, the total number of clients, the number of clients selected in each iteration, and the number of iterations will affect the allocation methods and amount added of DP noise and thereby affecting the model utility.
Research works in \cite{liu2020fedsel,truex2020ldp,wei2021user,kim2021federated} also design methods that adjust or reduce the amount of LDP noise by considering the above factors to improve the model usability.

\textbf{GDP.} 
Although LDP can protect the privacy of individual users in FL and can satisfy the situation where no trusted curator/server is available, it adds more noise and may cause more serious damage to model utility than GDP-based privacy-enhancing methods. Therefore, there are also some works that try to use GDP approaches to achieve PPAgg in FL.
The first work that considers GDP in FL optimization from the user-level is provided in \cite{geyer2017differentially}. The authors consider protecting the whole dataset of the clients instead of preserving only a single data entry by enhancing the privacy of the training models in FL. The central model $w_{t}$ is perturbed with Gaussian noise.
The authors in \cite{mcmahan2017learning} also apply GDP to secure the privacy of a deep language model on FedAvg and FedSGD algorithms. 
Both \cite{geyer2017differentially,mcmahan2017learning} show that the privacy of the trained FL model can achieve better accuracy when the number of local clients is large.
The proposed method in \cite{wei2020federated} perturbs local information by a GDP technique with Gaussian noise. They also provide a $K$-client random scheduling strategy to select users for FL model training.
In Noisy-FL \cite{chuanxin2020federated}, a privacy tracking framework f-DP \cite{dong2019gaussian} is leveraged to accurately track the privacy loss, and a GDP-based PPAgg protocol is employed on the global model to address the limitations that need a large number of clients in \cite{geyer2017differentially,mcmahan2017learning} with Gaussian mechanism.
The authors in \cite{hu2020personalized} design a personalized PPFL model in a heterogeneous IoT setting. Their model can achieve $(\epsilon, \delta)$-DP with $L_2$-sensitivity by adding the Gaussian noise to the global model during each iteration.
They assume that the dropped out users can rejoin the model training without disturbing the normal training process.

\textbf{Hybrid methods.}
Except for research works that only apply LDP-based or GDP-based protocols for secure aggregation in FL, there also exist hybrid approaches that combine LDP with GDP or other PPAgg techniques to both realize privacy protection of user data and prevention of model inversion.
User privacy and model privacy both are protected in \cite{xiong2021privacy} by leveraging LDP and GDP jointly.
The following are related works that combine DP-based protocols with other PP approaches.
The authors in \cite{hao2019efficient} apply additively HE and DP with the Gaussian mechanism to prevent privacy leakage from the local gradients and the shared models in FL, respectively. Their method can protect data privacy even in situations where the attacker colludes with multiple participants.
To achieve jointly tight record-level and user-level privacy guarantees, RDP and MPC are utilized in \cite{triastcyn2019federated}. 
Both local gradients and global models are protected. This method is under the assumption that the data distributions are similar among different users and the designed model is suitable for non-i.i.d (independent and identically distributed) data.
The work in \cite{wu2022adaptive} uses MPC and DP to make the client information indistinguishable and protect the global model update in FL.
LDP and function encryption are combined in \cite{yin2021privacy} to achieve both data-level and content-level privacy-preserving.
Compression and secure aggregation are combined with DP in \cite{andrew2021differentially} to guarantee both private and accurate models by an adaptive quantile clipping method.
LDP and Shuffled Model are utilized in \cite{ghazi2019scalable,girgis2021shuffled,sun2020ldp} to enhance model security by amplifying privacy through anonymization. 
An asynchronous model update scheme and a Malicious Node Detection Mechanism are designed to integrate with LDP in \cite{liu2021towards} for communication-efficient and attack-resistant Federated Edge Learning.
In \cite{agarwal2021skellam}, the Skellam mechanism instead of the Gaussian mechanism is introduced and the authors explore its performance when combining it with central RDP, distributed RDP with secure encryption, respectively.
The authors in \cite{hu2021concentrated} combine LDP with secure encryption and zCDP to achieve a good utility-privacy trade-off by adding less noise in every training iteration.
It is necessary to consider the problem of trading off the information privacy between model utility. The model utility includes but not limited to convergence performance, communication efficiency, and accuracy. 
Appropriately combining other PPAgg techniques with DP-based secure methods can improve the model utility.
Besides, choosing and allocating the privacy budget distribution should be carefully considered as it also has a significant effect on the utility of the FL trained models when using DP techniques. 


\subsection{Blockchain-based Aggregation}
\label{blockchain-based aggregation}

\begin{figure}[htbp!]
\vspace{-0.3cm}
    \centering
    \setlength{\abovecaptionskip}{1pt}
    \includegraphics[width=1\linewidth]{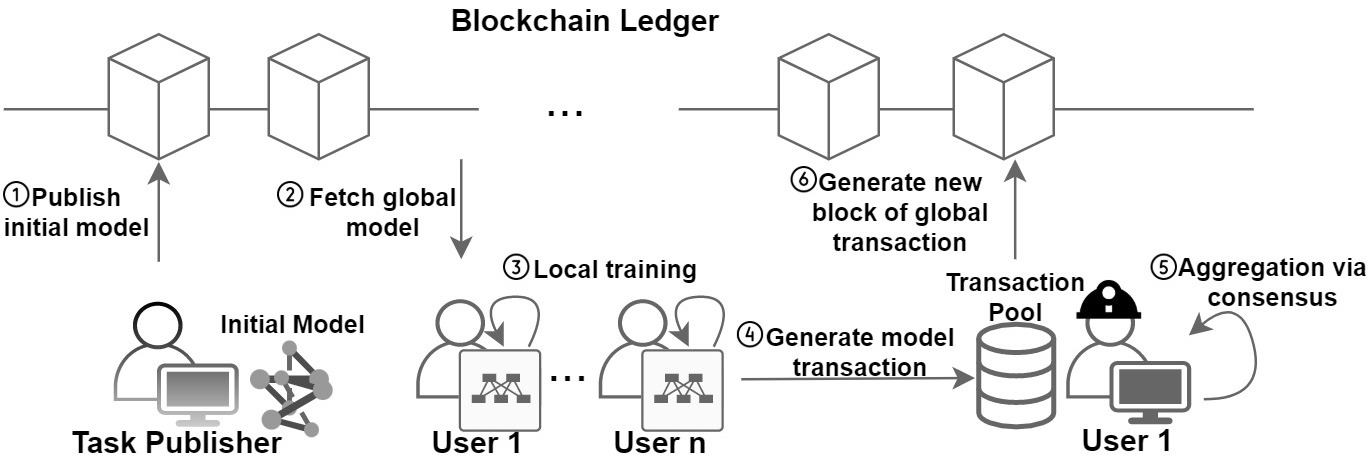}
    \caption{An illustrative figure of the aggregation in FL based on blockchain.}
    \label{fig:blockchain-agg}
\end{figure}

The nature of blockchain allows one to distribute the trust from a single server to a set of blockchain nodes, hence providing resilience to single-point-of-failure. In this case, a task publisher is usually involved to initialize the global FL model. Meanwhile, blockchain guarantees the auditability of the data and operation processed in the blockchain and the anonymity of participants (see Section \ref{sec:blockchain}). As shown in Figure \ref{fig:blockchain-agg}, a typical blockchain-based aggregation protocol in FL works as follows: (i) the task publisher publish FL training task with an initial global model to the blockchain, then (ii) each FL user fetches the global model, and (iii) trains its local model. After that (iv) each FL user generates and broadcasts a transaction recording its local model, which will be received and stored by nodes in their transaction pools, and then (v) the elected consensus node aggregate those local models to obtain the global model via the consensus protocol. Finally, (vi) the new global model is included in a block attaching to the blockchain for the next round of FL training.

\textbf{Integration of PPTs.} Note that the standard blockchain cannot prevent information leakage during model aggregation in FL, as users' models uploaded to blockchain are still over plaintext. Thus, additional PPTs are considered to be integrated to achieve the blockchain-based PPAgg protocol. For example, CDP methods are adopted in \cite{lu2019blockchain, zhao2020privacy, kim2020incentive, mondal2022beas} and LDP methods are adopted in \cite{chen2018machine, liu2020secure5g, ruckel2022fairness} to perturb users' models before uploading. Paillier crypto-systems are applied to \cite{bhagavan2021fedsmarteum,jia2021blockchain, zhu2021privacy} with its standard version as in the M1 setting, or with the threshold version \cite{weng2019deepchain} in the M3 setting (see Section \ref{sec:he-based aggregation}). However, security concerns still exist regarding the key management, i.e., arrangement on the ownership of secret keys used in adopted crypto-systems. Therefore, several studies aim to introduce a third party as an assistant. For example, in \cite{li2020crowdsfl}, the task publisher and the third party have the key-pair $(pk_1,sk_1)$ and $(pk_1,sk_1)$, respectively. They cooperate to generate a public key $pk_3$ that messages encrypted by $pk_3$ can be transformed, i.e., re-encrypted, to the one encrypted by $pk_1$ in some way using $sk_2$. In this case, FL users encrypt their models using the public key $pk_3$ and send them to the third party via blockchain, where re-encryption and aggregation are performed. Thus, security can be guaranteed as long as the task publisher and the third party is non-colluding. Apart from the above-mentioned works where users' models are uploaded or recorded by blockchain, several studies aim to leverage blockchain to store intermediate materials of existing PPAgg protocols for auditability. For example, \cite{jiang2021pflm} and \cite{fang2022privacy} rely on the SecAgg protocol \cite{bonawitz2017practical} where secret keys involved are shared and stored using blockchain. 

\textbf{Smart contract.} Beyond integrating PPTs to provide privacy guarantees upon blockchain, smart contract can be adopted to further enhance the security or efficiency of aggregation. For example, HE-based PPAgg protocol \cite{qi2021blockchain}, \cite{wu2020fedbc} and \cite{ majeed2021st} additionally leverage smart contract to generate the key-pair $(pk, sk)$ where the public key $pk$ is used to encrypt users' models while the secret key $sk$ is passed to a trusted party \cite{wang2021bpfl}, a leader elected by the blockchain consensus protocol \cite{qi2021blockchain}, or a key manager in smart contract \cite{wu2020fedbc,majeed2021st}. These schemes regarding key management are similar to those in the M1 setting discussed in Section \ref{sec:he-based aggregation}, however, the security risk is distributed to blockchain nodes, and the collusion risk is mitigated by using the smart contract.

\begin{table*}[htbp!]
\setlength{\abovecaptionskip}{2pt} 
\caption{A summary of PPAgg constructions. The provided security level, i.e., threat model, of the listed constructions is determined by their underlying protocols. Note that the ``large" and ``small" to describe the distribution scale correspond to those for cross-device and cross-silo FL settings.}
\label{tab:agg-comp}
\resizebox{\textwidth}{!}{%
\begin{tabular}{|c|c|c|c|c|c|c|c|c|c|c|} 
\hline
\multirow{2}{*}{\begin{tabular}[c]{@{}c@{}}\\\textbf{}\end{tabular}} & \multirow{2}{*}{\textbf{Masking}} & \multicolumn{3}{c|}{\textbf{HE}} & \multicolumn{2}{c|}{\textbf{MPC}} & \multicolumn{2}{c|}{\textbf{DP}} & \multirow{2}{*}{\textbf{Blockchain}} & \multirow{2}{*}{\textbf{TEE}} \\ 
\cline{3-9}
 &  & AHE & Threshold HE & FHE & Additive & Shamir & GDP & LDP &  &  \\ 
\hline
\begin{tabular}[c]{@{}c@{}}\textbf{Privacy }\\\textbf{ guarantee}\end{tabular} & \begin{tabular}[c]{@{}c@{}}Local \\ model\end{tabular} & \begin{tabular}[c]{@{}c@{}}Local \\ model\end{tabular} & \begin{tabular}[c]{@{}c@{}}Local \\ model\end{tabular} & \begin{tabular}[c]{@{}c@{}}Local model; \\ Global model\end{tabular} & \begin{tabular}[c]{@{}c@{}}Local model; \\ Global model\end{tabular} & Local model & \begin{tabular}[c]{@{}c@{}}Global \\ model\end{tabular} & \begin{tabular}[c]{@{}c@{}}Local \\ model\end{tabular} & N.A. & \begin{tabular}[c]{@{}c@{}}Local model; \\ Global model\end{tabular} \\ 
\hline
\begin{tabular}[c]{@{}c@{}}\textbf{Distribution }\\\textbf{ scale}\end{tabular} & Large & Large & Small & Small & Small & Large & Large & Small & Large & Small \\ 
\hline
\begin{tabular}[c]{@{}c@{}}\textbf{Resource }\\\textbf{ requirement}\end{tabular} & \begin{tabular}[c]{@{}c@{}}Usually \\ involves \\ interactions \\ among users\end{tabular} & Lightweight & \begin{tabular}[c]{@{}c@{}}Requires \\ expensive \\ key generation \\ and decryption\end{tabular} & \begin{tabular}[c]{@{}c@{}}Impractical \\ computation \\ overheads for \\ large-scale \\ ML models\end{tabular} & \begin{tabular}[c]{@{}c@{}}Usually involves \\ large \\ communication \\ overheads\end{tabular} & Lightweight & \multicolumn{2}{c|}{Negligible} & \begin{tabular}[c]{@{}c@{}}Costs of \\ transactions\end{tabular} & \begin{tabular}[c]{@{}c@{}}Requires the \\ deployment \\of TEE \\ hardware \\ platforms\end{tabular} \\ 
\hline
\begin{tabular}[c]{@{}c@{}}\textbf{Dropout-}\\\textbf{ resilience}\end{tabular} & Support & Support & \begin{tabular}[c]{@{}c@{}}Partially \\ support\end{tabular} & Support & NO & Support & \multicolumn{2}{c|}{Support} & Support & Support \\ 
\hline
\begin{tabular}[c]{@{}c@{}}\textbf{Model }\\\textbf{ utility}\end{tabular} & \multicolumn{6}{c|}{Depends on the encoding methods of underlying related cryptographic blocks} & \multicolumn{2}{c|}{Affected} & N.A. & N.A. \\ 
\hline
\begin{tabular}[c]{@{}c@{}}\textbf{Example }\\\textbf{ application}\end{tabular} & \cite{bonawitz2017practical,bell2020secure,liu2022efficient} & \cite{aono2017privacy,dong2020eastfly} & \cite{zheng2019helen,liu2020boosting}  & \cite{froelicher2021scalable,SavPTFBSH21}  & \cite{sharma2019secure,ryffel2018generic} & \cite{bonawitz2017practical,bell2020secure}  & \cite{mcmahan2017learning,wei2020federated} & \cite{ghazi2019scalable,zhao2020local} & \cite{zhao2020privacy,weng2019deepchain} & \cite{zhao2021sear,truex2020ldp} \\ 
\hline
\textbf{Remark} & \begin{tabular}[c]{@{}c@{}}Needs specific \\ desgin for \\ different FL \\ settings\end{tabular} & \multicolumn{2}{c|}{\begin{tabular}[c]{@{}c@{}}Key managements \\ affect security\end{tabular}} & \begin{tabular}[c]{@{}c@{}}Impractical for \\ deep neural \\ networks\end{tabular} & \begin{tabular}[c]{@{}c@{}}Suitable for \\only cross-silo \\FL settings\end{tabular} & \begin{tabular}[c]{@{}c@{}}Widely used \\as a building \\ block to \\ handle \\ dropped users\end{tabular} & \multicolumn{2}{c|}{\begin{tabular}[c]{@{}c@{}}Suffers the \\ loss of FL \\ model \\ performance\end{tabular}} & \begin{tabular}[c]{@{}c@{}}Needs to \\ integrate \\ with other \\ PPTs\end{tabular} & \begin{tabular}[c]{@{}c@{}}Limited \\ computation \\ resource and \\ costs of \\ hardware\end{tabular} \\
\hline
\end{tabular}
}
\end{table*}

\subsection{TEE-based Aggregation}
\begin{figure}[htbp!]
    \centering
    \setlength{\abovecaptionskip}{1pt}
    \includegraphics[width=0.65\linewidth]{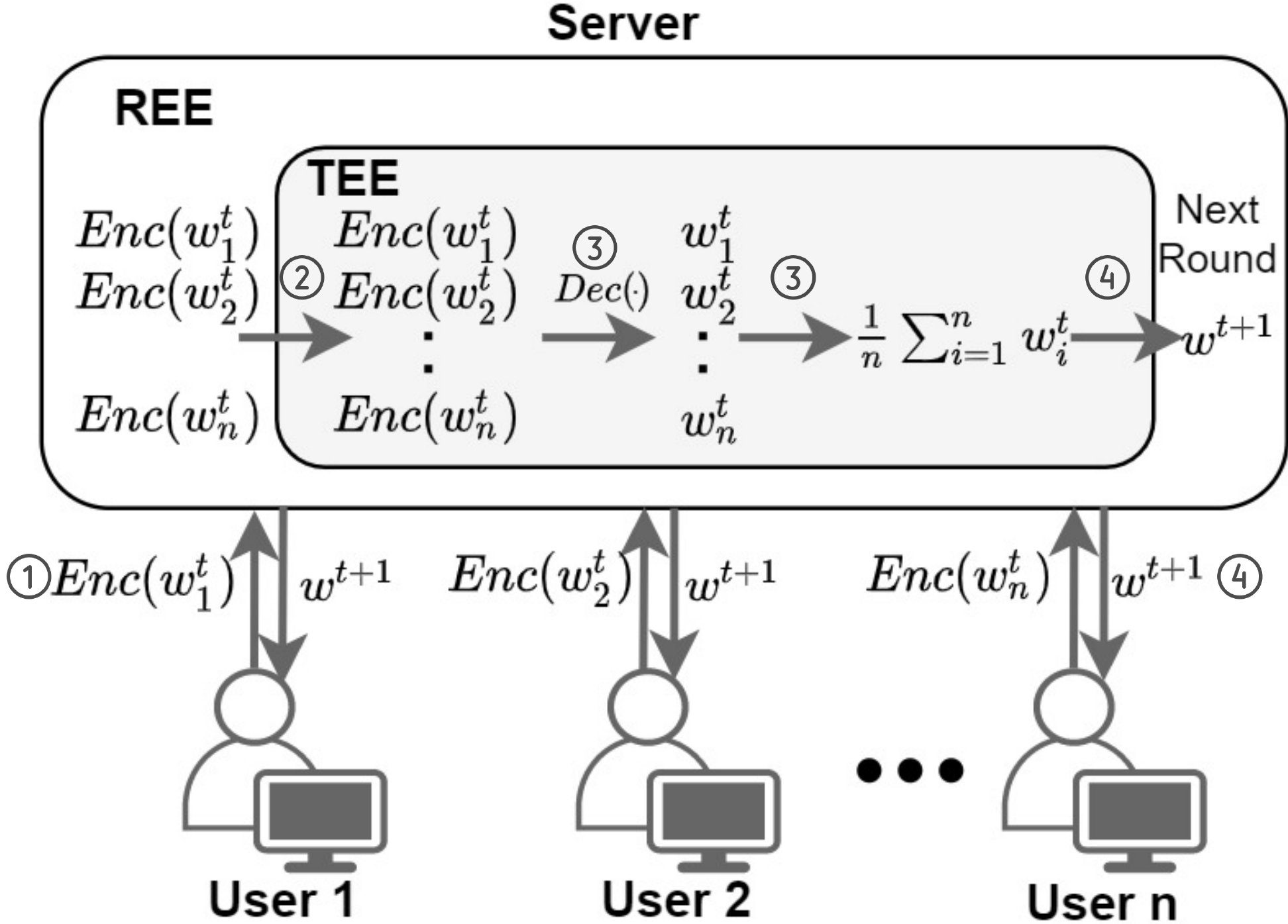}
    \caption{An illustrative figure of the aggregation in FL based on TEE.}
    \label{fig:reetee}
\end{figure}

As shown in Figure \ref{fig:reetee}, a typical TEE-based aggregation protocol in FL works as follows: (i) all users encrypt their locally trained models and send them to REE, (ii) TEE loads received encrypted models from REE, then (iii) decrypts and aggregates them. After that, (iv) TEE outputs the aggregation result to REE for the distribution to all users. The adoption of such typical TEE-based PPAgg can be found in \cite{mo2019efficient, hashemi2021byzantine, zhao2021sear}. To further enhance the security against potential attacks to TEE \cite{sabt2015trusted}, DP techniques are adopted to perturb users' models before uploading them to TEE \cite{mo2019efficient}. Alternatively, in \cite{zhang2021shufflefl}, the trust is distributed among several TEEs. In MixNN \cite{boutet2021mixnn}, a TEE is involved to shuffle users' models before uploading them to the central server, hence introducing randomness to enhance the security.

To extend TEE-based aggregation to the training process to further protect the privacy of the whole FL workflow, several studies aim to deploy ML training algorithms in TEE environments \cite{mo2019efficient, mo2020darknetz, kuznetsov2021securefl, mo2021ppfl,huang2021starfl}. However, the constrained memory size of the TEE platform usually leads to partial training in TEE \cite{mo2019efficient, mo2020darknetz, kuznetsov2021securefl}, or fully training in TEE with a long-time delay \cite{mo2021ppfl}.

Besides, signature and verification schemes can be integrated with TEE to provide integrity guarantees of honestly local training \cite{zhang2021tee,zhao2021sear,chen2020training}, or correct aggregation of the central server \cite{zheng2022aggregation,chen2020training}. However, we should note that as TEE-based PPAgg usually involves extra costs regarding hardware compared to the pure software-based PPAgg, large-scale deployment of TEEs in FL would be costly.

\subsection{Discussions}
\label{sec:discussion}

So far, we have reviewed PPAgg protocols with the discussions on the advantages and disadvantages of their constructions. Since FL systems are deployed under different settings with respect to the threat model, resource requirement, distribution scale, etc, we summarize the properties of these PPAgg constructions in Table~\ref{tab:agg-comp} as a reference for interested readers that aim to design their specific PPFL schemes.

In general, to provide privacy guarantees in an FL system, DP-based PPAgg is the most lightweight option but suffers from the loss of model performance. Thus, one has to adopt more expensive cryptographic tools, e.g., HE or MPC, to construct PPAgg to obtain the comparative exact solution of the target ML problem. Among those cryptographic tools, additive HE is the most natural option as it directly supports the homomorphic addition operation, i.e., aggregation. However, privacy issues exist due to the key management. In a standard HE scheme, neither the central server nor users keeping the secret key provide security against collusion. While threshold HE schemes that are with more robust key management inevitably involve expensive protocols of key generation and decryption. Besides, enabling the privacy protection of the whole FL workflow requires the adoption of FHE, which may lead to impractical overheads for large-scale ML models. Compared to HE, MPC is more efficient but still suffers from large communication overheads. Therefore, PPAgg protocols based on pure HE or MPC can be considered only at a cross-silo FL setting where participants have sufficient computation and communication capability and keep online from round to round. Compared to PPAgg based on pure cryptographic tools, masking-based PPAgg protocols are a more promising construction for large-scale PPFL schemes. They combine several lightweight cryptographic techniques, hence providing cost-effective execution for resource-constrained FL users. Besides, Shamir secret sharing schemes are usually adopted to handle dropped users. Therefore, masking-based PPAgg protocols are more suitable for a cross-device FL setting. However, the complicated constructions and lack of generality may hinder their wide deployment. To further improve the privacy and security of FL schemes, one can consider hybrid methods that integrate several PPTs, blockchain, or TEE to provide desired properties.

\begin{table*}[htb]
\scriptsize
\centering
\caption{Some open-source FL frameworks support PPAgg. A PPFL framework may consist of several PPAgg protocols for different settings, e.g., Horizontal FL (HFL) and Vertical FL (VFL), which provide different privacy guarantees. In general, the number of participants in VFL is less than that of HFL. Note that all listed frameworks provide privacy guarantee on users' locally trained modes and some of them support privacy-preserving training, hence protecting the privacy of the global model. If the PPAgg constructions provide security against active malicious settings, e.g., SPDZ and SecAgg, the framework also allows corresponding extensions.}
\label{tab:fl-framework}
\begin{tabular}{|c|c|c|c|c|l|}
\hline
\textbf{Framework} &
  \textbf{PPAgg construction} &
  \textbf{Privacy guarantee} &
  \textbf{Threat model} &
  \textbf{FL setting} &
  \multicolumn{1}{c|}{\textbf{Remark}} \\ \hline
TFF \cite{bonawitztensorflow} &
  Central DP \cite{abadi2016deep} &
  Global model &
  Semi-honest users &
  Cross-device &
  \begin{tabular}[c]{@{}l@{}}Partial protection of user model privacy; Available for \\only large-scale FL.\end{tabular} \\ \hline
FATE \cite{liu2021fate} &
  \begin{tabular}[c]{@{}c@{}}Paillier \cite{paillier1999public};  SPDZ \cite{damgaard2012multiparty}; \\ OT \cite{hauck2017efficient}; VSS \cite{feldman1987practical}; \\ SecAgg \cite{bonawitz2017practical}\end{tabular} &
  \begin{tabular}[c]{@{}c@{}}Local model; \\ Global model\end{tabular} &
  \begin{tabular}[c]{@{}c@{}}Semi-honest \\ users and server\end{tabular} &
  Cross-silo &
  \begin{tabular}[c]{@{}l@{}}Paillier and SecAgg for PPAgg protocols in HFL; SPDZ\\ and OT for PPAgg protocols in VFL.\end{tabular} \\ \hline
PaddleFL \cite{PaddleFL} &
  \begin{tabular}[c]{@{}c@{}}Central DP \cite{abadi2016deep}; \\SecAgg \cite{bonawitz2017practical}; ABY3 \cite{mohassel2018aby3}; \\PrivC \cite{he2019privc}\end{tabular} &
  \begin{tabular}[c]{@{}c@{}}Local model; \\ Global model\end{tabular} &
  \begin{tabular}[c]{@{}c@{}}Semi-honest \\ users and server\end{tabular} &
  Cross-silo &
  \begin{tabular}[c]{@{}l@{}}Central DP and SecAgg for PPAgg protocols in HFL; \\Generic MPC protocols, i.e., ABY3 for MPC and PrivC \\for 2PC, for PPAgg protocols in VFL.\end{tabular} \\ \hline
  PySyft \cite{pysyft} &
  \begin{tabular}[c]{@{}c@{}}Central DP \cite{abadi2016deep}; \\ SPDZ \cite{damgaard2012multiparty};  CKKS \cite{cheon2017homomorphic}; \\ Paillier \cite{paillier1999public}\end{tabular} &
  \begin{tabular}[c]{@{}c@{}}Local model; \\ Global model\end{tabular} &
  \begin{tabular}[c]{@{}c@{}}Semi-honest \\ users and server\end{tabular} &
  \begin{tabular}[c]{@{}c@{}}Cross-silo; \\ Cross-device\end{tabular} &
  \begin{tabular}[c]{@{}l@{}}Integration with PyGrid API for the FL mode; Support-\\ing specific deployments on Android and iOS.\end{tabular} \\ \hline
Flower \cite{beutel2020flower} &
  \begin{tabular}[c]{@{}c@{}}SecAgg \cite{bonawitz2017practical}; \\ SecAgg+ \cite{bell2020secure}\end{tabular} &
  Local model &
  \begin{tabular}[c]{@{}c@{}}Semi-honest \\ users and server\end{tabular} &
  \begin{tabular}[c]{@{}c@{}}Cross-silo; \\ Cross-device\end{tabular} &
  \begin{tabular}[c]{@{}l@{}}Mainly designed for large-scale FL settings with hetero-\\geneous participants.\end{tabular} \\ \hline
\end{tabular}
\end{table*}

\section{Federated Learning Frameworks for Privacy-Preserving Aggregation}
\label{sec:fl-frameworks}

With the active development of federated learning, many FL frameworks have been proposed as open-source libraries to support follow-up works and to enable easy deployment and replicability of FL systems. In Table \ref{tab:fl-framework}, we list some existing open-source FL frameworks that support privacy-preserving aggregation with a summary of the construction of PPAgg protocols, privacy guarantees, and threat models.

Note that in addition to the PPFL frameworks listed in Table \ref{tab:fl-framework}, there are also PPFL frameworks under development from some leading IT companies and organizations, e.g., FedML \cite{he2020fedml}, PrivacyFL\cite{mugunthan2020privacyfl}, HyFed \cite{nasirigerdeh2021hyfed}, Federated Learning and Differential Privacy framework by Sherpa.ai \cite{sherpa}, Hive by Ping An Technology \cite{hive_pingan}, Fedlearn-Algo by JD Finance \cite{liu2021fedlearn}, Huawei Noah's Ark FL framework \cite{chen2018federated}, and some distributed with proprietary or limited licenses, e.g, the FL framework by Ant Group \cite{fang2021large}, NVIDIA Clara \cite{clara_nvidia} and IBM-FL \cite{ludwig2020ibm}.

\section{Challenges and Future Directions}
\label{sec:challenges}

In Section \ref{sec:ppa-protocols}, we provide an in-depth survey on applications of PPAgg protocols to address a wide range of privacy and security issues in FL systems. However, with the fast evolution of PPFL schemes and their deployments, a plethora of emerging problems remain open for further studies, many of which require new PPAgg protocols to provide additional properties and support more operations. In this section, we expand our discussion to some challenges as well as research directions with PPFL systems, where PPAgg protocols may exert their further potential.

\subsection{Throughput Improvement}

The privacy-preserving aggregation protocols have been adopted in a lot of PPFL schemes. However, the throughput, i.e., the capacity of processing aggregation operations, of many PPAgg protocols limits the scope of PPFL applications, especially for large-scale networks. The reason is that their building blocks leverage generic expensive cryptographic techniques, which involve large computation or communication overheads. Thus, specific lightweight cryptographic techniques designed for aggregation in FL are required, e.g., communication-efficient masking-based algorithm \cite{bell2020secure} and lightweight additive HE (AHE) scheme \cite{jiang2021flashe}. Besides, as machine learning algorithms typically consist of a large number of vector operations, efficient integration of batch operations should be considered to improve the throughput, e.g., batch encryption \cite{ZhangLX00020}, SIMD techniques in HE schemes \cite{cheon2017homomorphic}, and parallelized hardware architectures such as FPGA and GPU. Furthermore, the combination of compression techniques with efficient PPAgg protocols can greatly reduce the communication overheads in FL but remains a challenge. This is because compression techniques, e.g., Top-k sparsification \cite{aji2017sparse}, require the order information to reconstruct the original vector from the compressed vector. Such information vector may lead to severe privacy leakage, while reconstruction over encrypted order information vector is not straightforward. Therefore, proper integration of different techniques with PPAgg protocols for throughput improvement can be further investigated.


\subsection{Hybrid Schemes for Stronger Security}

In addition to the privacy threats discussed in this survey, other attacks from a security perspective may also hinder the deployment of FL systems, e.g., poisoning attacks \cite{huang2011adversarial} and inference attacks \cite{shokri2017membership}. Thus, integrating PPAgg protocols with other security algorithms to construct hybrid schemes can be considered for future research.

\textbf{Poisoning attack.} Poisoning attacks aim to reduce the accuracy of the FL model, i.e., random attacks \cite{huang2011adversarial}, or induce the FL model to output the target label specified by the adversaries, i.e., targeted attacks or backdoor attacks \cite{bagdasaryan2020backdoor}, by manipulating local data or models. Therefore, PPAgg protocols cannot provide security against poisoning attacks. For poisoning attacks from the user-side, the central server can adopt Byzantine-resilient aggregation algorithms, e.g., \cite{yin2018byzantine,blanchard2017machine,CaoF0G21} to detect anomalies. However, these schemes require access to the users' data or models which violates the privacy goal of PPFL, hence cannot be integrated with PPAgg protocols. Therefore, only Byzantine-resilient aggregation algorithms that do not access users' data or models such as \cite{andreina2021baffle} or those that work over encrypted data can be adopted for integration, which remains for further investigation. For poisoning attacks from the server-side, one needs to guarantee that the server correctly aggregates the models from users. TEE, blockchain, verifiable secret sharing (VSS), and verifiable computation techniques can be applied to PPAgg protocols for the verification.

\textbf{Inference attack.} Inference attacks aim to cause information leakages of users' data, e.g., membership \cite{shokri2017membership}, attribute \cite{melis2019exploiting}, or data \cite{zhu2019deep}, by probing a target ML model. Most PPAgg protocols mitigate this issue by protecting the users' models. However, privacy leakages still exist in some PPFL systems with PPAgg protocols. For example, as pointed out in \cite{naseri2020local}, LDP-based aggregation does not guarantee security against attribute inference attacks, while GDP-based aggregation works only when sacrificing significant utility. Besides, in many PPFL systems, global models are revealed to adversaries, which are still vulnerable to inference attacks. Therefore, integrating other privacy-preserving techniques with PPAgg protocols to enhance security remains a topic for further research.

\section{Conclusions}
\label{sec:conclusions}

This paper has presented a comprehensive survey of the privacy-preserving aggregation protocols adopted to enhance the privacy of federated learning systems. Firstly, we have given an overview of federated learning on its concepts, data organization, working mechanism, and privacy threats to FL systems. Then, we have introduced the basic knowledge of supporting tools for constructing PPAgg protocols. Afterward, we have provided reviews and analyses of different constructions of PPAgg protocols in detail to deal with a variety of privacy issues in FL systems. Finally, we have outlined existing challenges as well as several directions for future research.

\ifCLASSOPTIONcaptionsoff
  \newpage
\fi

\bibliographystyle{IEEEtran}
\bibliography{bare_jrnl_compsoc}

\vspace{-1cm}
\begin{IEEEbiography}[{\includegraphics[width=1in,height=1.25in,clip,keepaspectratio]{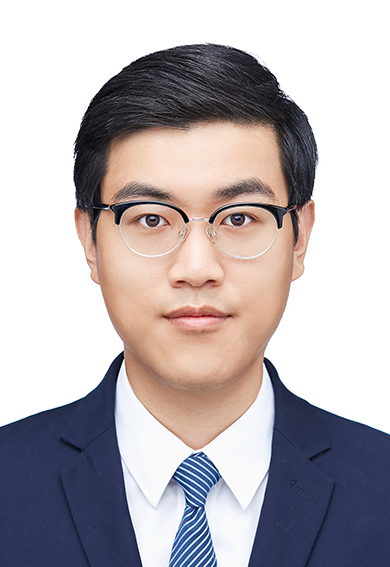}}]{Ziyao Liu}
received his B.E. degree from the school of Electronics Information Engineering, Zhengzhou University, Zhengzhou, China, in 2015, and the M.S. degree from Beijing Institute of Technology, Beijing, China, in 2018. He is currently working towards a Ph.D. degree in the School of Computer Science and Engineering, Nanyang Technological University, Singapore. His research interests include privacy-preserving machine learning, multi-party computation, and applied cryptography.
\end{IEEEbiography}
\vspace{-1cm}

\begin{IEEEbiography} [{\includegraphics[width=1in,height=1.25in,clip,keepaspectratio]{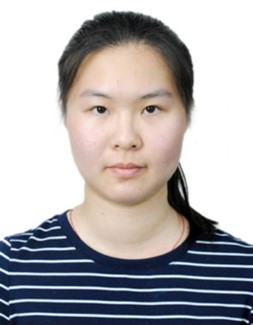}}]{Jiale Guo}
received her B.S. from the School of Mathematics, Shandon University, Jinan, China, in 2017. She is currently pursuing a Ph.D. degree in the School of Computer Science and Engineering, Nanyang Technological University, Singapore. Her research interests include Privacy-Preserving machine learning and Cybersecurity. 
\end{IEEEbiography}
\vspace{-1cm}
\begin{IEEEbiography}[{\includegraphics[width=1in,height=1.25in,clip,keepaspectratio]{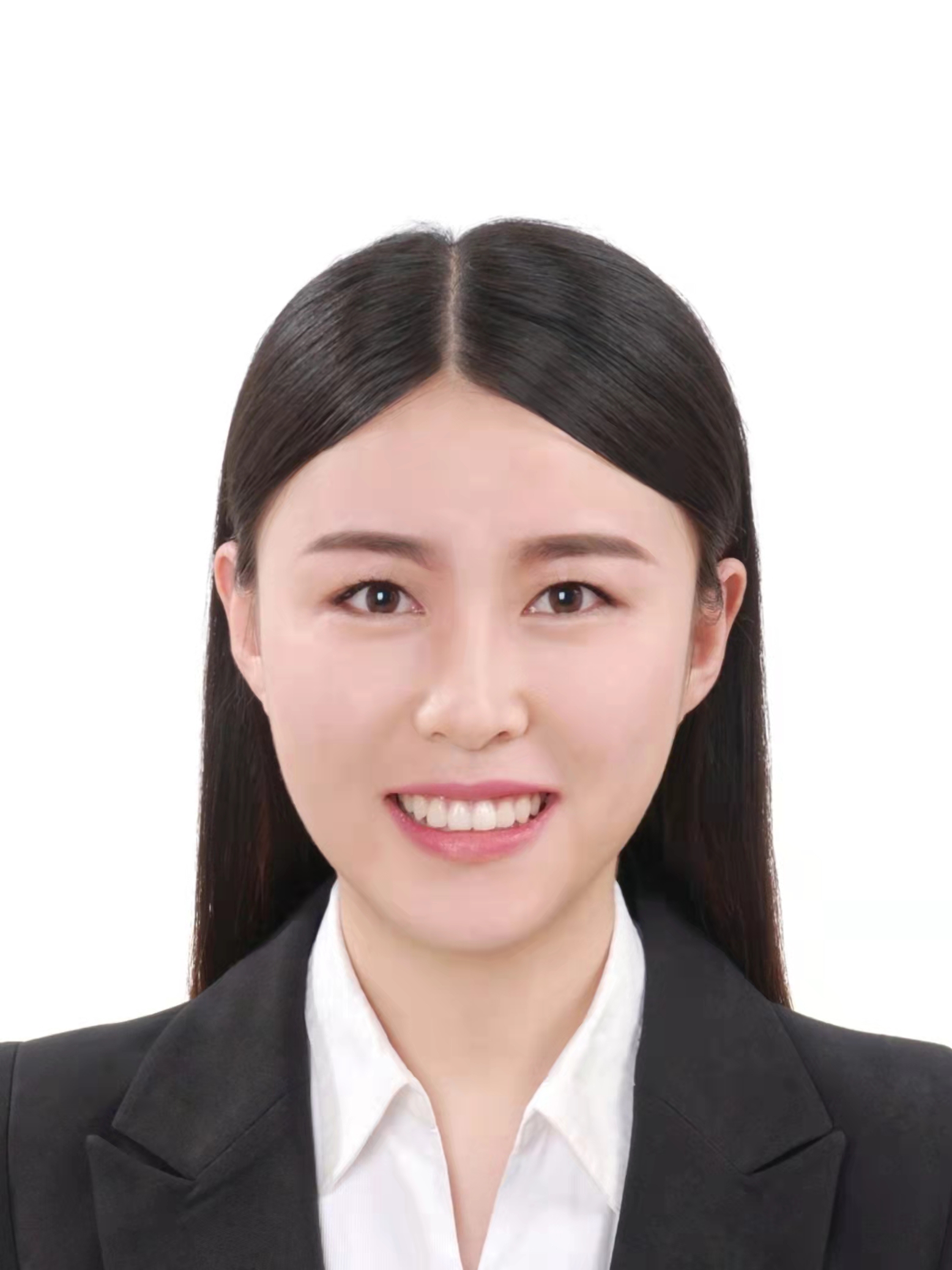}}]{Wenzhuo Yang} received the Bachelor’s degree in Measuring and Controlling Technologies and Instruments from Beijing University of Posts and Telecommunications, Beijing, China, in 2016. She is currently pursuing the Ph.D. degree with the School of Computer Science and Engineering, Nanyang Technological University, Singapore. Her current research interests are in the area of Privacy-Preserving Machine Learning, IoT Security, Intrusion Detection, and Cyber Threat Intelligence Analysis.
\end{IEEEbiography}
\vspace{-1cm}
\begin{IEEEbiography}[{\includegraphics[width=1in,height=1.25in,clip,keepaspectratio]{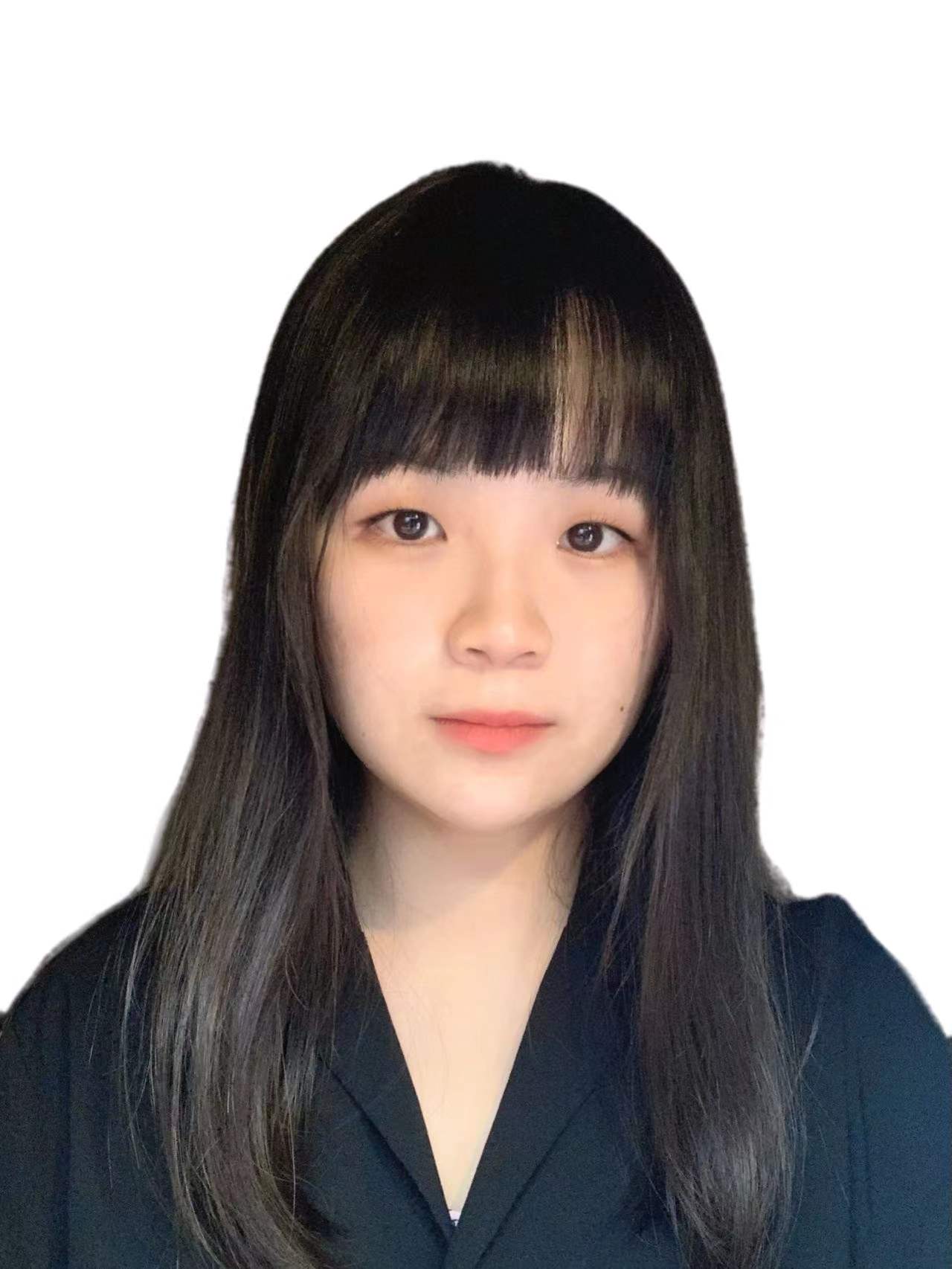}}]{Jiani Fan} received the Bachelor’s degree in information systems from Singapore Management University in 2020. She is currently pursuing the Ph.D. degree in computer science at Nanyang Technological University. Her research interests include IoT Security, Cybersecurity, and Internet of Vehicles.
\end{IEEEbiography}
\vspace{-1cm}

\begin{IEEEbiography} [{\includegraphics[width=1in,height=1.25in,clip,keepaspectratio]{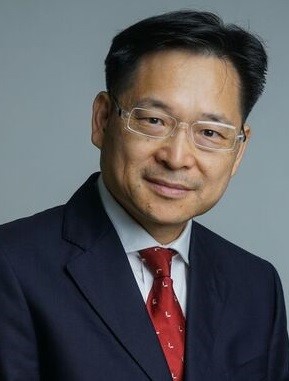}}]{Kwok-Yan Lam}
(Senior Member, IEEE) received his B.Sc. degree (1st Class Hons.) from University of London, in 1987, and Ph.D. degree from University of Cambridge, in 1990. He was a Visiting Scientist at the Isaac Newton Institute, Cambridge University, and a Visiting Professor at the European Institute for Systems Security. He has collaborated extensively with law-enforcement agencies, government regulators, telecommunication operators, and financial institutions in various aspects of Infocomm and Cyber Security in the region. From 2002 to 2010, he was a Professor with Tsinghua University, China. Since 1990, he has been a Faculty Member with the National University of Singapore and the University of London. He is currently a Full Professor with Nanyang Technological University, Singapore and the Director of the Strategic Centre for Research in Privacy-Preserving Technologies and Systems (SCRiPTS). From August 2020, Professor Lam is also on part-time secondment to the INTERPOL as a Consultant at Cyber and New Technology Innovation. In 1998, he received the Singapore Foundation Award from the Japanese Chamber of Commerce and Industry in recognition of his research and development achievement in information security in Singapore.
\end{IEEEbiography}
\vspace{-1cm}
\begin{IEEEbiography} [{\includegraphics[width=1in,height=1.25in,clip,keepaspectratio]{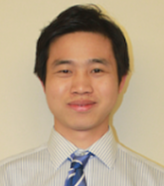}}]{Jun Zhao}
(S'10-M'15) is currently an Assistant Professor in the School of Computer Science and Engineering (SCSE) at Nanyang Technological University (NTU), Singapore. He received a Ph.D. degree in Electrical and Computer Engineering from Carnegie Mellon University (CMU), Pittsburgh, PA, USA, in May 2015, and a bachelor's degree in Information Engineering from Shanghai Jiao Tong University, China, in June 2010. One of his papers was a finalist for the best student paper award in IEEE International Symposium on Information Theory (ISIT) 2014. His research interests include A.I. and data science, security and privacy, control and learning in communications and networks.
\end{IEEEbiography}

\end{document}